\documentclass[aps,pra,twocolumn,showpacs,superscriptaddress,floatfix,nofootinbib]{revtex4}
\usepackage{graphicx}
\usepackage{amsmath}
\usepackage{epsfig}
\usepackage{helvet}
\usepackage{amssymb}

\newcommand{\tmop}[1]{\ensuremath{\operatorname{#1}}}

\newcommand{\tr}{\mbox{tr}}
\newcommand{\bra}[1]{\mbox{$\langle #1 |$}}
\newcommand{\ket}[1]{\mbox{$| #1 \rangle$}}
\newcommand{\braket}[2]{\mbox{$\langle #1  | #2 \rangle$}}
\newcommand{\proj}[1]{\mbox{$|#1\rangle \!\langle #1 |$}}

\begin{document}

\title{Simulation of two-dimensional quantum systems using a tree tensor network that exploits the entropic area law.}
\author{L. Tagliacozzo}
\author{G. Evenbly}
\author{G. Vidal}
\affiliation{School of Physical Sciences, the University of
Queensland, QLD 4072, Australia} 
\date{\today}

\begin{abstract}
This work explores the use of a tree tensor network ansatz to simulate the ground state of a local Hamiltonian on a two-dimensional lattice. By exploiting the entropic area law, the tree tensor network ansatz seems to produce quasi-exact results in systems with sizes well beyond the reach of exact diagonalisation techniques. We describe an  algorithm to approximate the ground state of a local Hamiltonian on a $L \times L$ lattice with the topology of a torus. Accurate results are obtained for  $L=\{4,6,8\}$, whereas approximate results are obtained for larger lattices. As an application of the approach, we analyse the scaling of the ground state entanglement entropy at the quantum critical point of the model.  We confirm the presence of a positive additive constant to the area law for half a torus.  We also find a  logarithmic additive correction to the entropic area law for a square block.  The single copy entanglement for half a torus reveals similar corrections to the area law with a further term proportional to  $1/L$.
\end{abstract}

\maketitle


\section{Introduction}

The numerical study of many-body quantum systems is a challenging task. The exponential growth of the dimension of the Hilbert space with the size of the system implies that exact diagonalisation techniques can only be applied to address small lattice systems \cite{hamer_finite-size_2000,henkel_statistical_1984,laeuchli_dynamical_2009}. Quantum Monte Carlo sampling offers a valuable route to the study of larger lattices, although systems of frustrated quantum spins or interacting fermions cannot be analysed due to the so called \emph{sign problem}.

In two spatial dimensions, the use of a tensor network ansatz, such as the tensor product state or projected entangled pair state (PEPS)\cite{verstraete_renormalization_2004,murg_variational_2007,jordan_classical_2008,murg_exploring_2009,sierra_density_1998, maeshima_vertical_2001,nishio_tensor_2004,gu_tensor-entanglement_2008} and the multi-scale entanglement renormalisation ansatz (MERA) \cite{vidal_entanglement_2007,vidal_class_2006,evenbly_entanglement_2008,evenbly_algorithms_2007}, has opened a very promising alternative path to investigating ground state properties of arbitrarily large lattice systems. The key of these approaches is the ability of the TPS,  PEPS and MERA to reproduce the scaling of the entanglement in the ground state, as given by the entropic area law.

In this work we explore the use of yet another tensor network variational ansatz, namely a \emph{tree tensor network} (TTN) \cite{fannes_ground_1992,niggemann_quantum_1997,friedman_density_1997,lepetit_density-matrix_2000,martn-delgado_density-matrix_2002,shi_classical_2006,nagaj_quantum_2008, otsuka_density-matrix_1996}, to simulate the ground state of local 2D lattice systems. This very simple ansatz is inspired on the original real space Renormalisation Group ideas of Kadanoff, Migdal and Wilson \cite{Kadanoff:1966wm,kadanoff_static_1967,wilson_renormalization_1975,fisher_renormalization_1998,burkhardt_real-space_1982,kadanoff_variational_1975}.

The present approach is both motivated and limited by the  area law for the entanglement entropy \cite{bombelli_quantum_1986,srednicki_entropy_1993,plenio_entropy_2005,masanes_area_2009,latorre_ground_2003,eisert_area_2008}. The "area law" is a conjectured property of the entanglement entropy
of  certain ground states of local  Hamiltonians. It asserts  that the
entropy of a region of the  system is proportional to
the size of its boundary rather than proportional to its volume.
Direct calculations of the entanglement entropy have provided evidence
of the validity of the area law for a large class of systems \cite{bombelli_quantum_1986,srednicki_entropy_1993,masanes_area_2009,plenio_entropy_2005}.
A complete characterization  of the Hamiltonians whose  ground state obeys
 the area law is still missing.

On the one hand, by exploiting the area law a TTN can be used to address small 2D lattices with sizes well beyond the reach of exact diagonalisation techniques. Specifically, the cost of simulating a lattice of $L\times L$ sites grows as $\exp(L)$ instead of $\exp(L^2)$. Thus, the TTN approach is useful to investigate small 2D quantum systems and to study larger systems with finite size scaling techniques. It is also particularly suitable to investigate ground state entropies.

On the other hand, the $\exp(L)$ cost due to the entropic area law still sets a severe limit to the system sizes a TTN can describe and the present approach simply cannot compete with the  PEPS and MERA algorithms \cite{nishio_tensor_2004,gu_tensor-entanglement_2008,verstraete_renormalization_2004,jordan_classical_2008,evenbly_entanglement_2008} for large systems. However, the TTN is also of interest in the context of developing these more advanced, scalable algorithms. This is due both to its simplicity and to its direct connection to ground state entanglement properties, on which the scalable algorithms are also based. As a matter of fact, the TTN approach described in this work was initially developed as an auxiliary tool to help in the design of the MERA \cite{evenbly_entanglement_2008}.

The present approach bears important similarities with White's \emph{density matrix renormalisation group} (DMRG) \cite{white_real-space_1992,white_density_1992,noack_real-space_1993} (for a review see e.g.  \cite{schollwck_density-matrix_2005}) when applied to 2D lattice systems \cite{shibata_application_2003,white_ne[e-acute]l_2007,white_density_1998,liang_approximate_1994,henelius_two-dimensional_1999,du_croo_de_jongh_critical_1998,xiang_two-dimensional_2001}. Roughly speaking, it can be regarded as a DMRG approach where the \emph{matrix product state} has been replaced with a TTN. This replacement has both advantages and disadvantages. Its weakest point is an increase in computational cost. However, a TTN greatly improves the connectivity between lattice sites, possibly resulting in faster convergence and better correlation functions (e.g. on a torus). Extracting certain entropies from the TTN, say the entropy of one quarter of the lattice, is straightforward. Finally, the algorithm can be very simply implemented.

 The results are organised in several sections. In Sect. \ref{sect:ansatz} we describe the TTN for 2D lattices and motivate its use in terms of the entropic area law. In Sect. \ref{sect:computation} we explain how to compute the expectation value of local operators, two-point correlation functions, fidelities and block entropies. Then in Sect. \ref{sect:algorithm} we describe an algorithm to approximate ground states with a TTN. This algorithm is tested in Sect. \ref{sect:benchmark} by addressing the quantum Ising model with transverse magnetic field on a torus made of $L\times L$ sites. Quasi exact results are obtained for lattices of linear size $L=\{4,6,8\}$, whereas approximate results are obtained for $L=\{10,16,32\}$, we also check  the exponential  cost of a TTN representation and introduce a possible estimate of the error induced by the finite amount of computational resources.

In Sect. \ref{sect:entropy} we turn our attention to the computation of ground state entanglement at the critical point. We present results for $L=\{4,6,8,10\}$ for both entanglement entropy and single copy entanglement. This allows us to investigate the form of their finite size  scaling.  Several authors \cite{casini_universal_2007,fradkin_entanglement_2006,hsu_universal_2008,stphan_shannon_2009,fradkin_scaling_2009,yu_entanglement_2008,casini_entanglement_2009,metlitski_entanglement_2009,casini_entanglement_2005,nishioka_holographic_2009}  have predicted the presence of corrections to the area law including a universal and positive constant  term for half a torus \cite{fradkin_scaling_2009, metlitski_entanglement_2009,stphan_shannon_2009}.
 Our results reproduce this term, as well as a logarithmic correction for a quarter of a torus \cite{fradkin_entanglement_2006,yu_entanglement_2008,casini_entanglement_2005,nishioka_holographic_2009}.
%
Numerical estimates of the coefficients for  all the corrections are presented in Eqs. \ref{eq:num_ris_s}, \ref{eq:num_ris_s1}, \ref{eq:num_ris_l}, \ref{eq:num_ris_e}-\ref{eq:num_ris_e3}.  We conclude with a discussion of the results in Sect. \ref{sect:discussion}.

\begin{figure}[!htb]
\begin{center}
\includegraphics[width=8cm]{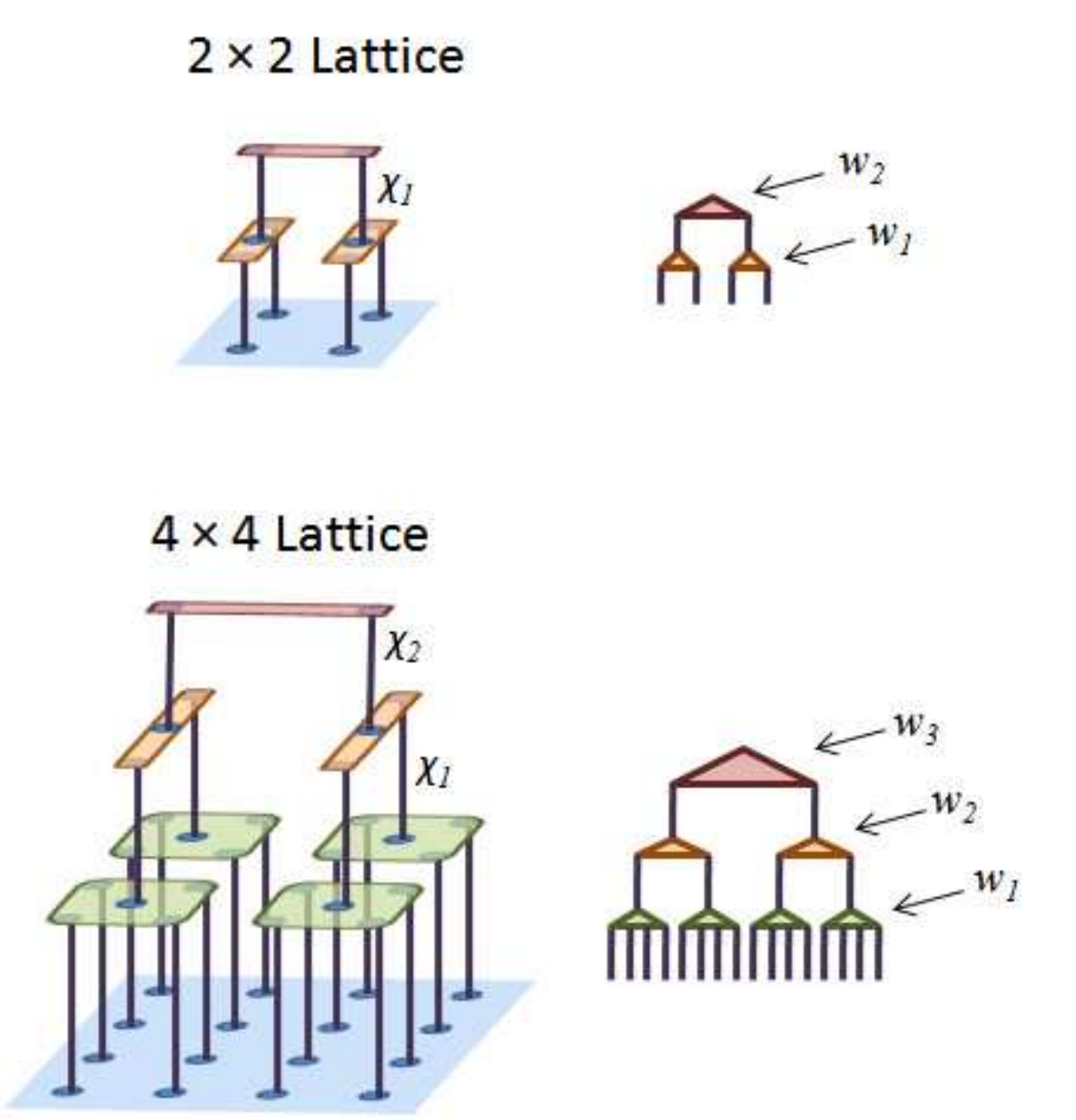}

\caption{Example of TTN for a $2\times 2$ lattice and a $4\times 4$ lattice. Notice (right) that the TTN for a 2D lattice can always be represented as a planar graph, with the leaves or physical indices ordered on a line. {The tensors  labelled with $w_i$ are isometric tensors. They act locally by  projecting  the ground  state onto its local support with dimension $\chi_i$ } (see section \ref{sect:ansatz} for further explanation).} 
\label{fig:SmallTree}
\end{center}
\end{figure}

\section{Tree Tensor Network Ansatz}
\label{sect:ansatz}

In this section we introduce the variational ansatz used throughout the manuscript and justify its applicability in terms of the area law for entanglement entropy.

\subsection{Isometric tree tensor network}

We consider a square lattice $\mathcal{L}$ made of $N = L\times L$ sites, where each site is described by a local Hilbert space $\mathbb{V}$ of finite dimension $d$. Our goal is to represent a pure state $\ket{\Psi} \in \mathbb{V}^{\otimes N}$ of the lattice $\mathcal{L}$. Most of the time, $\ket{\Psi}$ will correspond to the ground state $\ket{\Psi_{\mbox{\tiny GS}}}$ of some local Hamiltonian $H$ defined on $\mathcal{L}$.

A generic state $\ket{\Psi}\in \mathbb{V}^{\otimes N}$ can always be expanded as
\begin{eqnarray}
	\ket{\Psi} = \!\sum_{i_1=1}^d ~ \sum_{i_2=1}^d \cdots \sum_{i_N=1}^d T_{i_1i_2 \cdots i_N} \ket{i_1} \ket{ i_2} \cdots \ket{i_N},
\label{eq:local_expansion}
\end{eqnarray}
where the $d^{N}$ coefficients $T_{i_1i_2 \cdots i_N}$ are complex numbers and the vectors $\{\ket{1_s}, \ket{2_s}, \cdots, \ket{d_s}\}$ denote a local basis on site $s\in \mathcal{L}$. We refer to the index $i_s$ that labels a local basis for site $s$ ($i_s=1,\cdots,d$) as a \emph{physical} index. 

In this work we further expand the tensor of coefficients $T_{i_1i_2 \cdots i_N}$ in Eq. \ref{eq:local_expansion} using a TTN. As shown in Fig. \ref{fig:SmallTree} for lattices of $2\times 2$ and $4\times 4$ sites, a TTN decomposition consists of a collection of tensors $w$ that have both \emph{bond} indices and \emph{physical} indices. The tensors are interconnected by the bond indices according to a tree pattern. The $N$ physical indices correspond to the leaves of the tree. Upon summing over all the bond indices, the TTN produces the $d^N$ complex coefficients $T_{i_1i_2 \cdots i_N}$ of Eq. \ref{eq:local_expansion}. 

\begin{figure}[!tb]
\begin{center}
\includegraphics[width=6cm]{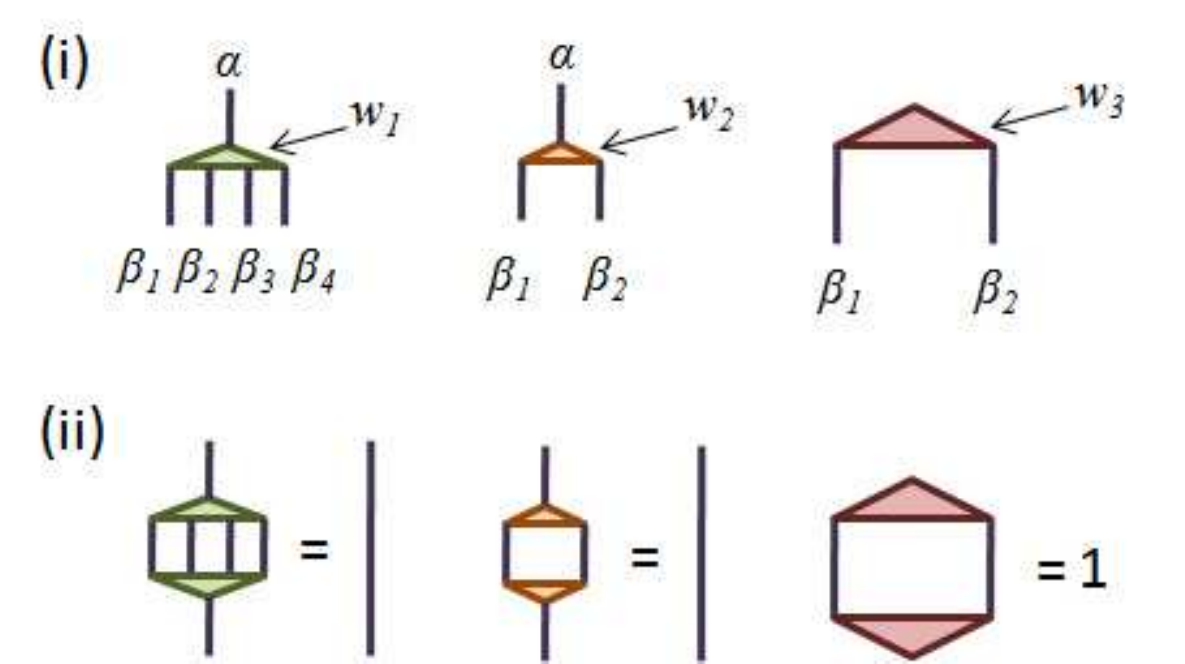}
\caption{ (i) Diagrammatic representation of three types of isometric tensors in the TTN for a $4\times 4$ lattice in Fig. \ref{fig:SmallTree}. (ii) Graphical representation of the constraints in Eqs. \ref{eq:const1}-\ref{eq:const3} fulfilled by the isometric tensors. } 
\label{fig:Isometric}
\end{center}
\end{figure}

The tensors in the TTN will be constrained to be \emph{isometric}, in the following sense. As shown in Fig. \ref{fig:Isometric} for the $4\times 4$ case of Fig. \ref{fig:SmallTree}, each tensor $w$ in a TTN has at most one upper leg/index $\alpha$ and some number $p$ of lower indices/legs $\beta_1,\cdots, \beta_p$, so that its entries read $(w)^{\alpha}_{\beta_1\cdots\beta_p}$. Then we impose that
\begin{eqnarray}
	\sum_{\beta_1 \cdots \beta_p} (w)_{\beta_1 \cdots \beta_p}^{\alpha}(w^{\dagger})^{\beta_1 \cdots \beta_p}_{\alpha'} = \delta_{\alpha\alpha'}.
	\label{eq:isometry}
\end{eqnarray}
For the sake of clarity, throughout the paper we use diagrams to represent tensor networks as well as tensor manipulations. For instance, the constraints for the tensors $w_1$, $w_2$ and $w_3$ of the TTN of Fig. \ref{fig:SmallTree} for a $4\times 4$ lattice, namely
\begin{eqnarray}
	\sum_{\beta_1 \beta_2 \beta_3 \beta_4} (w_1)_{\beta_1 \beta_2 \beta_3 \beta_4}^{\alpha} (w_1^{\dagger})^{\beta_1 \beta_2 \beta_3 \beta_4}_{\alpha'} &=& \delta_{\alpha\alpha'}, \label{eq:const1}\\
	\sum_{\beta_1 \beta_2} (w_2)_{\beta_1 \beta_2}^{\alpha}(w_2^{\dagger})^{\beta_1 \beta_2}_{\alpha'} &=& \delta_{\alpha\alpha'}, \\
		\sum_{\beta_1 \beta_2} (w_3)_{\beta_1 \beta_2}(w_3^{\dagger})^{\beta_1 \beta_2} &=& 1,\label{eq:const3}
\end{eqnarray}
are represented as diagrams in Fig. \ref{fig:Isometric}(ii). We refer to a tensor $w$ that fulfils Eq. \ref{eq:isometry} as an \emph{isometry}. As we will see in Sects. \ref{sect:computation} and \ref{sect:algorithm}, the use of isometries simplifies the manipulations necessary to compute sxpectation values of local operators and the spectrum of reduced density matrices, as well as to optimise the TTN. The isometric character of the tensors can also be seen to prevent numerical instability during the simulations.

\subsection{Coarse-graining of the lattice}

\begin{figure}[!tb]
\begin{center}
\includegraphics[width=8cm]{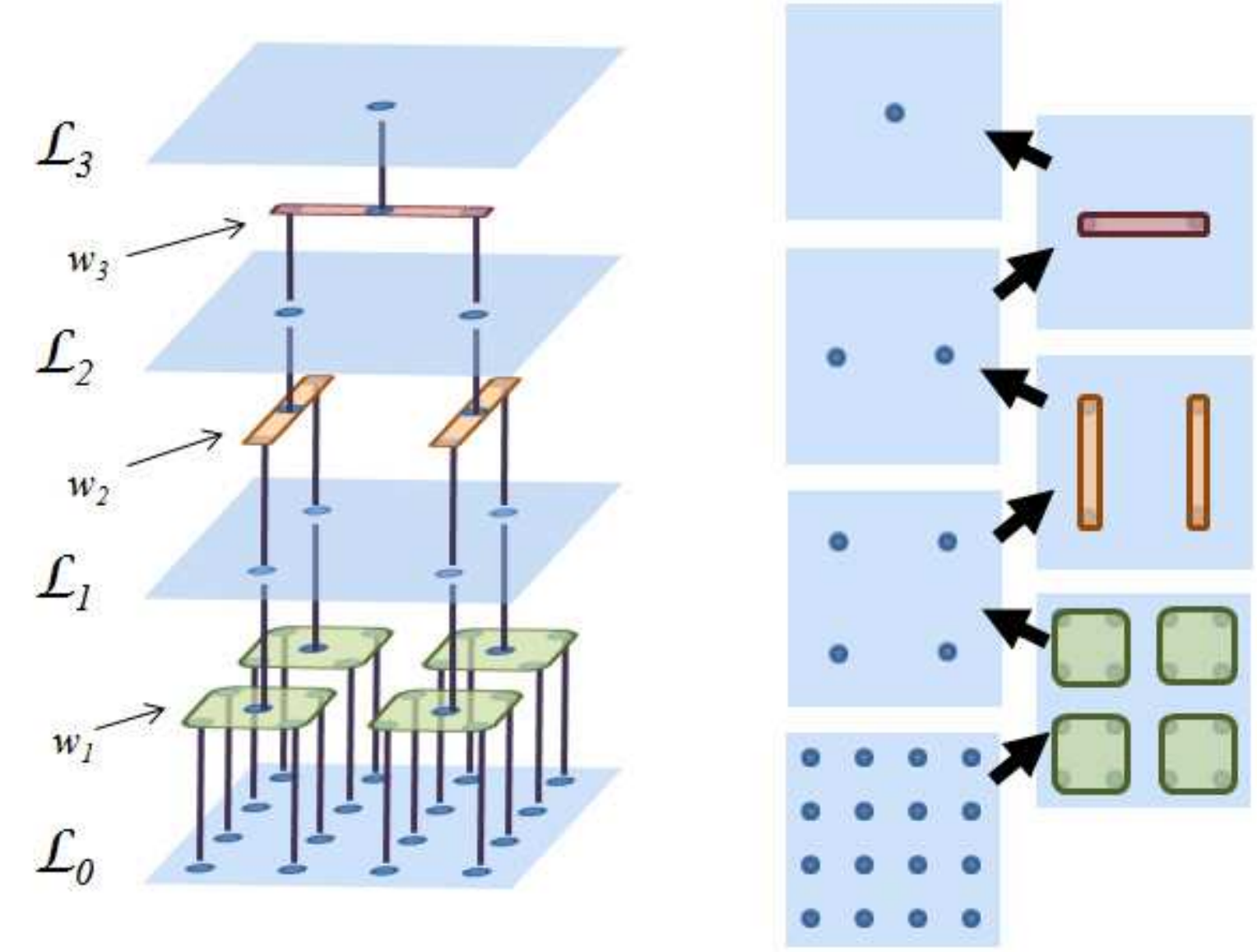}
\caption{ The isometric TTN of Fig. \ref{fig:SmallTree} for a $4\times 4$ lattice $\mathcal{L}_0$ is associated with a coarse-graining transformation that generates a sequence of increasingly coarse-grained lattices $\mathcal{L}_1$, $\mathcal{L}_2$ and $\mathcal{L}_3$. Notice that in this example we have added an extra index to the top isometry $w_{3}$, corresponding to the single site of an extra top lattice $\mathcal{L}_3$, which we can use to encode in the TTN a whole subspace of $\mathbb{V}^{\otimes N}$ instead of a single state $\ket{\Psi}$.} 
\label{fig:CoarseGrain}
\end{center}
\end{figure}

An intuitive interpretation of the use of a TTN to represent a state $\ket{\Psi}$ can be obtained in terms of a coarse-graining transformation for the lattice $\mathcal{L}$. Notice that the isometries $w$ in Fig. \ref{fig:SmallTree} are organised in layers. The bond indices between two layers can be interpreted as defining the sites of an effective lattice. In other words, the TTN defines a sequence of increasingly coarser lattices $\{\mathcal{L}_0, \mathcal{L}_1, \cdots, \mathcal{L}_{T-1} \}$, where $\mathcal{L}_0 \equiv \mathcal{L}$ and each site of lattice $\mathcal{L}_{\tau}$ is defined in terms of several sites of $\mathcal{L}_{\tau-1}$ by means of an isometry $w_{\tau}$, see Fig. \ref{fig:CoarseGrain}. 

In this picture, a site of the lattice $\mathcal{L}_{\tau}$ effectively corresponds to some number $n_{\tau}$ of sites of the original lattice $\mathcal{L}_0$. For instance, each of the two sites of $\mathcal{L}_{2}$ in Fig. \ref{fig:CoarseGrain} corresponds to $8$ sites of $\mathcal{L}_0$. Similarly, each site of lattice $\mathcal{L}_{1}$ corresponds to $4$ sites of $\mathcal{L}_0$.

\subsection{Entropic area law}
\label{subs:area_law}
In using a TTN to represent a generic state  $\ket{\Psi}$, the top tensor $w_{T}$ must already depend on $d^N$ coefficients. It is then unclear that the use of   a TTN has any computational advantage with respect to directly dealing with  all the $d^N$ coefficients $T_{i_1 i_2 \cdots i_N}$ in Eq. \ref{eq:local_expansion}. However, ground states $\ket{\Psi_{\mbox{\tiny GS}}}$ of local Hamiltonians are known to often exhibit a so-called entropic \emph{area law} \cite{bombelli_quantum_1986,srednicki_entropy_1993,plenio_entropy_2005,masanes_area_2009,latorre_ground_2003,eisert_area_2008} and this property might lead to a reduction in computational costs when expressing the state (or an accurate approximation of it) as a TTN.

Let us introduce the reduced density matrix $\rho$ for a block $A$ of contiguous sites of $\mathcal{L}$ as
\begin{equation}
	\rho^A = \tr_{B} \proj{\Psi} = \sum_{\alpha} p_{\alpha} \proj{\Psi^{A}_{\alpha}},
	\label{eq:rhoA}
\end{equation}
where $B$ are all the sites of $\mathcal{L}$ outside the block $A$ and $p_{\alpha}$ are the eigenvalues of $\rho^A$ (that is, $\rho^A\ket{\Psi^{A}_\alpha} = p_{\alpha} \ket{\Psi^{A}_{\alpha}}$). Then the entropy $S(A)$ of block $A$ is defined as
\begin{equation}
S(A) \equiv -\tr(\rho^A \log \rho^A) = -\sum_{\alpha} p_{\alpha} \log p_{\alpha}.
\label{eq:entropy}
\end{equation}
This entropy measures the amount of entanglement between the block $A$ and the rest $B$ of the lattice $\mathcal{L}$, and it is also known as \emph{entanglement entropy} \cite{bennett_concentrating_1996}. For a generic state, the entropy of a block $A$ is proportional to the number $n(A)$ of sites in $A$ (provided $n(A) \leq N/2$), that is
\begin{equation}
	S(A) \approx n(A) \log d ~~~~~~ \mbox{(generic)}
\end{equation}
For instance, the entropy of a block made of $l\times l$ sites is proportional to $l^2$ and, correspondingly, the effective dimension $\chi$ required to describe the block
\footnote{Throughout the paper we use a number of different subscripts to denote different effective dimensions $\chi$. For instance, $\chi_{l\times l}$, $\chi(A)$ and $\chi_{1/2}$ refer, respectively, to the effective dimension for a block of $l\times l$ sites, a block $A$ and one half of the lattice. The specific meaning should be clear from the context.}
is exponential in $l^2$,  
\begin{eqnarray}
	S_{l\times l} \approx l^2\log d,~~~~~~\chi_{l\times l} = d^{l^2} ~~~~ \mbox{(generic)}
\end{eqnarray}
 
{If instead the entropy of the block  $A$ grows proportional to the number of sites of the boundary of $A$, denoted by  $\Sigma(A)$,   we say that the state $\ket{\Psi}$ fulfils an entropic "area law"}
\begin{equation}
	S(A) \approx c_1 \Sigma(A)~~~~~~ \mbox{(area law)}
\label{eq:area_law}
\end{equation}
where $c_1$ is some constant. For instance, for the above block of $l\times l$ sites, the entropy is only proportional to $l$. Accordingly, the dimension $\chi$ required to effectively describe the block may grow markedly less with $l$ than in the generic case,
\begin{equation}
	S_{l\times l} \approx 4c_1 l, ~~~~~~ \chi_{l\times l} \geq \exp{(4c_1 l)}, ~~~~ \mbox{(area law)}
	\label{eq:lxl_area_law}
\end{equation}
where the lower bound for $\chi$ is obtained by exponentiating the entropy and is saturated by a flat probability distribution $p_{\alpha} = 1/\chi$,  $\alpha=1,\cdots,\chi$.

Eq. \ref{eq:lxl_area_law} is our main justification for attempting to describe ground states of local $2 D$  Hamiltonians using a TTN. It suggests that it might be possible to accurately approximate a ground state $\ket{\Psi_{\mbox{\tiny GS}}}$ that fulfils the area law of Eq. \ref{eq:area_law} by using a number of coefficients that scales with the linear size $L$ of the lattice $\mathcal{L}$ only as $O( \exp(L))$, instead of $O(\exp(L^2))$ as is the case for a generic state. In other words, ground states of local Hamiltonians are typically less entangled than generic states, and we might be able to exploit this fact computationally.

A simple example of ground state that fullfills the boundary law is a valence bond crystal {(as e.g. it is the ground state of the AKLT model \cite{affleck_rigorous_1987})} made of short-range dimers. Each dimer crossing the boundaries of a region $A$  contributes a constant amount to the entropy and the number of such dimers is proportional to the size of the boundary of the region $A$. This roughly corresponds to the presence, in the spectrum of the reduced density matrix, of a number of significant eigenvalues that grows exponentially with the size of the boundary. 

\subsection{Plane, cylinder and torus}

Let us now assume that the entropic boundary law in Eq. \ref{eq:area_law} translates into an effective site dimension given by
\begin{equation}
	\chi(A) \approx \exp(S(A)) \approx \exp(c_1 \sigma(A)),
\label{eq:chi_area_law}
\end{equation}
and let us explore the implications that this expression would have on the ability of a TTN to encode ground states. 

\begin{figure}[!tb]
\begin{center}
\includegraphics[width=8cm]{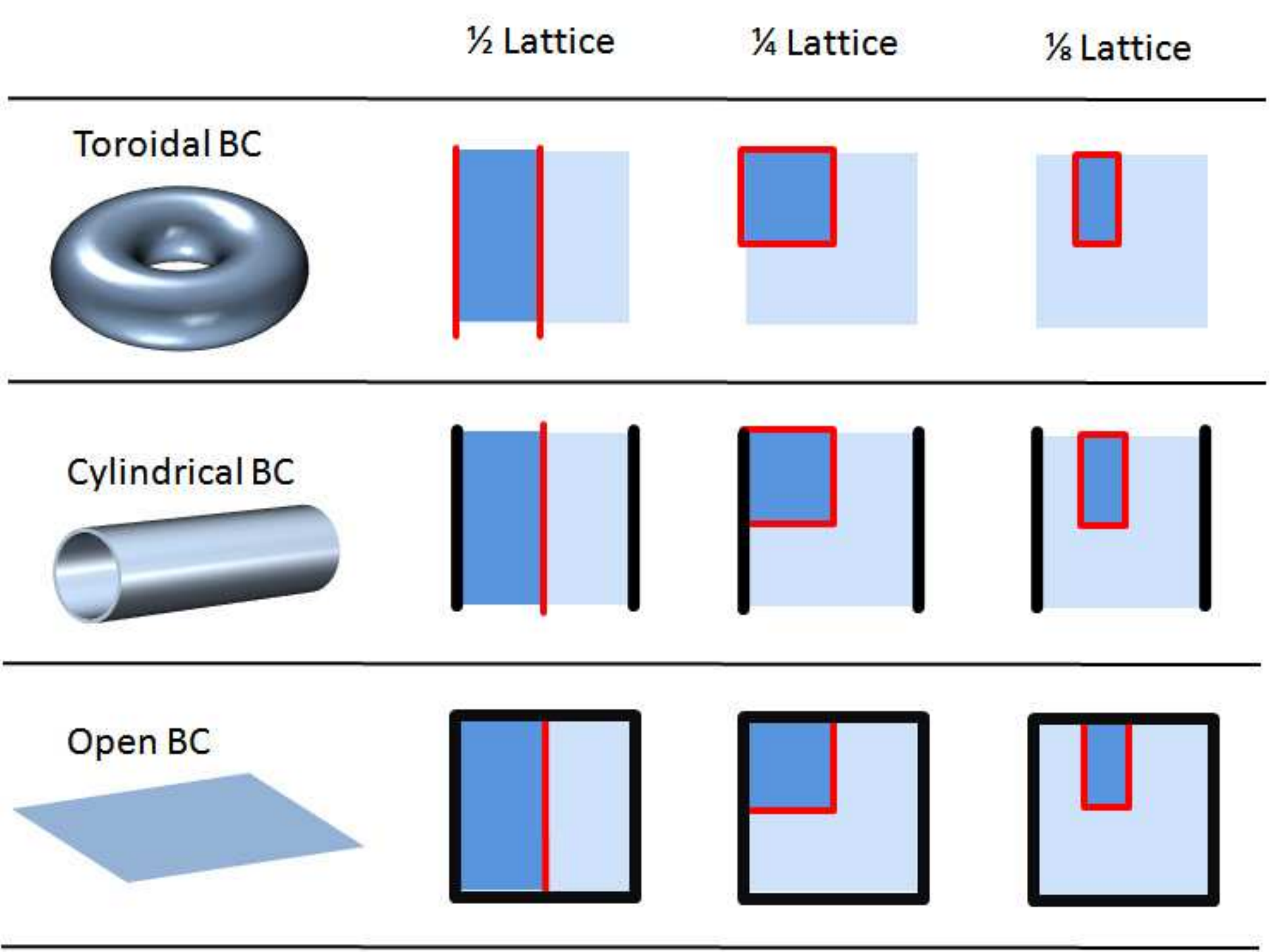}
\caption{Interacting boundaries $\Sigma_{1/2}$, $\Sigma_{1/4}$ and $\Sigma_{1/8}$ corresponding to one half, one quarter and one eighth of a lattice for three different choices of boundary conditions.} 
\label{fig:Boundary}
\end{center}
\end{figure}

For this purpose, let us consider the (interacting) boundaries, denoted $\Sigma_{1/2}$, $\Sigma_{1/4}$ and $\Sigma_{1/8}$, of blocks that consists, respectively, of one half, one fourth and one eighth of a $L\times L$ lattice $\mathcal{L}$. These boundaries depend on the topology of the interactions of $H$ on $\mathcal{L}$, and for the plane, cylinder and torus are  presented in table \ref{table:bound} (see also Fig. \ref{fig:Boundary}).
\begin{table}
\begin{equation}
\begin{tabular}{|c|c|c|c|}
  \hline
                & plane    &  cylinder & torus \\ \hline
 $\Sigma_{1/2}$ & L        & L        & 2L \\
 $\Sigma_{1/4}$ & L        & $\frac{3}{2}L$        & 2L \\
 $\Sigma_{1/8}$ & $\frac{5}{4}L$      & $\frac{3}{2}L$    & $\frac{3}{2}L$ \\
  \hline
  
\end{tabular} \nonumber
\end{equation}
\caption{{Length of the boundaries of different subregions of a 2D lattice system for several choices of topology of the whole system.} \label{table:bound}}
\end{table}

From  table \ref{table:bound}  and Eq. \ref{eq:chi_area_law} one can obtain the dimension $\chi$ of the sites of the most coarse-grained lattices $\mathcal{L}_{T\!-1}$, $\mathcal{L}_{T\!-2}$, $\mathcal{L}_{T\!-3}$, and the size of the isometries at the upper layers of the TTN, which is what dominates the computational cost of the approach. The table shows that ground states on a torus are more entangled (e.g. the blocks have more interacting boundary, or entropy), and therefore computationally more demanding, than on a plane or cylinder. In this work we shall concentrate on the torus, with the understanding that a similar analysis can also be conducted for the other cases. [In particular, as it is easy to anticipate, given the same computational costs, larger systems can be addressed in the cases of plane and cylinder interaction topologies.] 

\subsection{TTN ansatz on the torus}

From now on we consider a $L\times L$ lattice $\mathcal{L}$ on the torus. In this case, $\chi_{1/2} \approx \chi_{1/4} \approx \exp(c_1 2 L)$ are the largest effective site dimensions. The top isometry $w_{T}$ depends on $\chi_{1/2}^2 \approx \exp(4c_1L)$ parameters, whereas each isometry $w_{T-1}$ depends on $\chi_{1/2}\chi_{1/4}^2 \approx \exp(6c_1L)$ parameters. Isometries at lower layers of the TTN can be seen to depend on less parameters. 

\begin{figure}[!tb]
\begin{center}
\includegraphics[width=8cm]{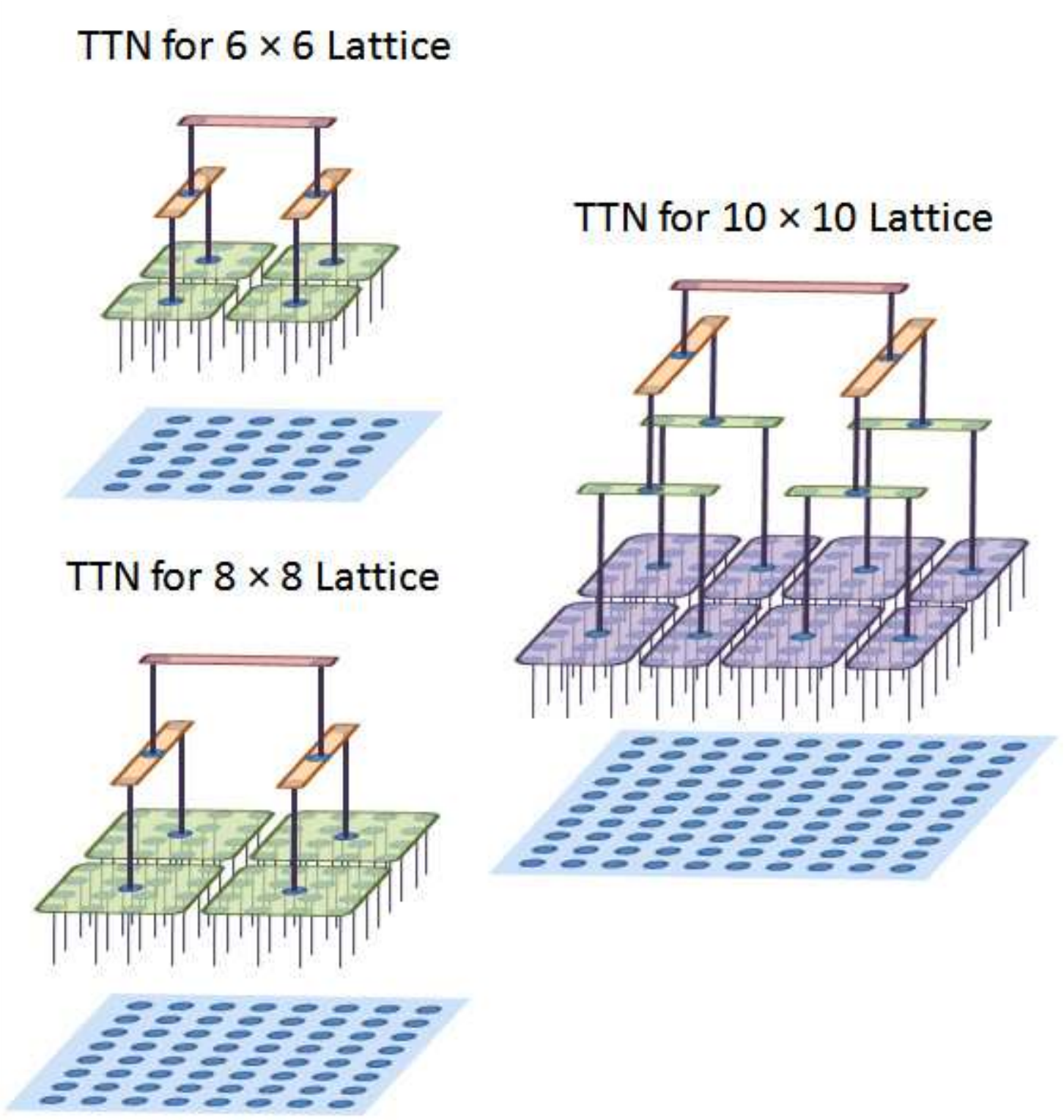}
\caption{ Isometric TTN for lattices of $6\times 6$, $8\times 8$ and $10 \times 10$ sites as used in the manuscript for the purpose of benchmarking the performance of the algorithm of Sect. \ref{sect:algorithm}. Notice that all TTN have the same structure on the two top layers of isometries, whose manipulation dominates the computational cost of the algorithm, while they differ in the lower layers. In particular, in the $10\times 10$ lattice two lower layers of isometries are required, since a single layer of isometries mapping a block of $5\times 5$ sites directly into a single effective site would have been too expensive given the present capabilities of the desktop computers used for the simulations.} 
\label{fig:BigTree}
\end{center}
\end{figure}

Based on these observations, our TTN ansatz for the ground state of an $L\times L$ lattice with torus topology and site dimension $d=2$ (e.g. spin-$\frac{1}{2}$ model) will invariably consists of a top isometry $w_T$ and two isometries $w_{T-1}$ with bond dimension $\chi$ on all their indices. Then, depending on the size $L$ and other considerations, the TTN will be completed in two possible ways. For small $L$ ($L\leq 8$ in the examples of section \ref{sect:benchmark}), a single extra layer of isometries will be considered, where each isometry maps $N/4$ sites of $\mathcal{L}$ directly into one site of $\mathcal{L}_{T-2}$. For larger lattices, it is computationally favourable to complete the TTN with at least two more layers of isometries, see Fig. \ref{fig:BigTree}.

Because the isometries $w_{T-1}$ are, by far,  the largest tensors in the TTN, the memory required to store the ansatz is a function of the size of $w_{T-1}$, namely
\begin{equation}
	\mbox{Memory } \approx  \chi^3,~~~~~~~~\mbox{(large $\chi$ regime)}
	\label{eq:memory}
\end{equation}
where, unless otherwise specified, from now on $\chi$ refers to the effective dimension used for both one half and one quarter of the lattice, $\chi \equiv \chi_{1/2} = \chi_{1/4}$.

\subsection{Nested Schmidt decompositions}

It is instructive to relate the TTN ansatz with the Schmidt decomposition of the state $\ket{\Psi}$ it represents. 

Recall that given a bipartition $A:B$ of the sites of lattice $\mathcal{L}$ into two subsets $A$ and $B$, the Schmidt decomposition of state $\ket{\Psi}$ according to this bipartition reads
\begin{equation}
	\ket{\Psi} = \sum_{\alpha=1}^{\chi(A:B)} \sqrt{p_{\alpha}} \ket{\Psi_{\alpha}^{A}}\ket{\Psi_{\alpha}^{B}},
\end{equation}
where $p_{\alpha}$, $\ket{\Psi^A_{\alpha}}$ and $\ket{\Psi^B_{\alpha}}$ appear in the spectral decomposition of the reduced density matrices (cf. Eq. \ref{eq:rhoA}) 
\begin{eqnarray}
	\rho^A = \sum_{\alpha} p_{\alpha} \proj{\Psi^{A}_{\alpha}},~~
	\rho^B = \sum_{\alpha} p_{\alpha} \proj{\Psi^{B}_{\alpha}},
\end{eqnarray}
and where the number of terms $\chi(A:B)$ in the decomposition, known as the Schmidt rank, can be used as a measure of entanglement between blocks $A$ and $B$ \cite{vidal_efficient_2003}.

In Ref. \cite{shi_classical_2006} a canonical form for the TTN was proposed, where each bond index of the TTN corresponds to a Schmidt decomposition. That is, in its canonical form, a TTN can be regarded as a collection of Schmidt decompositions of a state according to a family of nested bipartitions $A:B$ of the system.

In this work we do not use the canonical form of a TTN. However, the use of isometric tensors implies that the rank of each bond index in our TTN is given by the Schmidt rank $\chi(A:B)$ of the corresponding partition. In particular, the bond dimension $\chi$ in Eq. \ref{eq:memory} corresponds to the Schmidt rank between two halves of the system, as well as between one fourth and three fourths of the system.

\subsection{Symmetries}

The symmetries of a state $\ket{\Psi}$ of the lattice $\mathcal{L}$ can often be incorporated to some extent into the TTN, resulting in a reduction on computational costs. One can distinguish between \emph{space symmetries}, such as invariance under translations e.g. by one lattice site or invariance under rotation of the lattice by e.g. 90$^\circ$, and \emph{internal symmetries}, such as particle number conservation or spin isotropy.

The coarse-graining implicit in the TTN ansatz is incompatible with most space symmetries. As a result, a TTN approximation to a symmetric state typically breaks such symmetries. However, the symmetry is seen to be restored in the limit of a large $\chi$. In addition, the isometries can often explicitly incorporate part of the symmetry. For instance, in approximating  states that are invariant under translations in $4\times 4$, $6\times 6$ or $8\times 8$ lattices by using the TTNs of Figs. \ref{fig:SmallTree} and \ref{fig:BigTree}, one can choose all the isometries on a given layer of the TTN to be the same.

In contrast, internal symmetries can be implemented exactly in the TTN. Suppose for example that the state is known to have a well defined particle number (U(1) symmetry) or to be a singlet under spin rotations (SU(2) symmetry). Then one can choose all the isometries of the tree to be covariant under the action of the symmetry, in such a way that: ($i$) the symmetry is preserved \emph{exactly} by any value of $\chi$, and ($ii$) many parameters of the isometries are fixed by the symmetry, leading to a significant reduction in computational cost. We refer to \cite{singh_tensor_2009,mcculloch_non-abelian_2002,singh_matrix_2007} for more details.
{In the actual computations presented in this work we have not implemented internal  symmetries in the TTN.}
\subsection{Relation to real-space RG}

Being based on coarse-graining the lattice $\mathcal{L}$, the present approach is closely related to the real-space RG ideas and methods proposed by Kadanoff, Migdal and Wilson \cite{Kadanoff:1966wm,kadanoff_static_1967,wilson_renormalization_1975,fisher_renormalization_1998,burkhardt_real-space_1982,kadanoff_variational_1975}. The TTN ansatz can indeed be regarded as a specific implementation of the spin-blocking schemes that these authors put forward.

However, it is important to emphasise the differences between the present approach and those usually associated to real-space RG methods. First of all, here we attempt to obtain a quasi-exact description of the ground state $\ket{\Psi_{\mbox{\tiny GS}}}$ of a finite lattice $\mathcal{L}$, which forces us to consider effective sites with a dimension $\chi_{\tau}$ that grows (exponentially!) with the number of iterations $\tau$ of the coarse-graining transformation. Instead, real-space RG approaches typically attempt to identify and characterise the fixed points of the RG flow on an infinite system and consider a fixed dimension $\chi_{\tau}$. A second important difference is in the way the isometries are chosen. {Wilson proposed to consider the restriction $H_{B}$ of the Hamiltonian $H$ on a block of sites $B$, and to choose the isometries in order to preserve the subspace corresponding to the lowest energy eigenvalues of $H_{B}$. Here, instead, we aim at globally minimising $H$
 (see Sect. \ref{sect:algorithm}), thereby following the path initiated with White's density matrix renormalisation group (DMRG)} \cite{white_real-space_1992,white_density_1992,noack_real-space_1993}. 

\begin{figure}[!tb]
\begin{center}
\includegraphics[width=6cm]{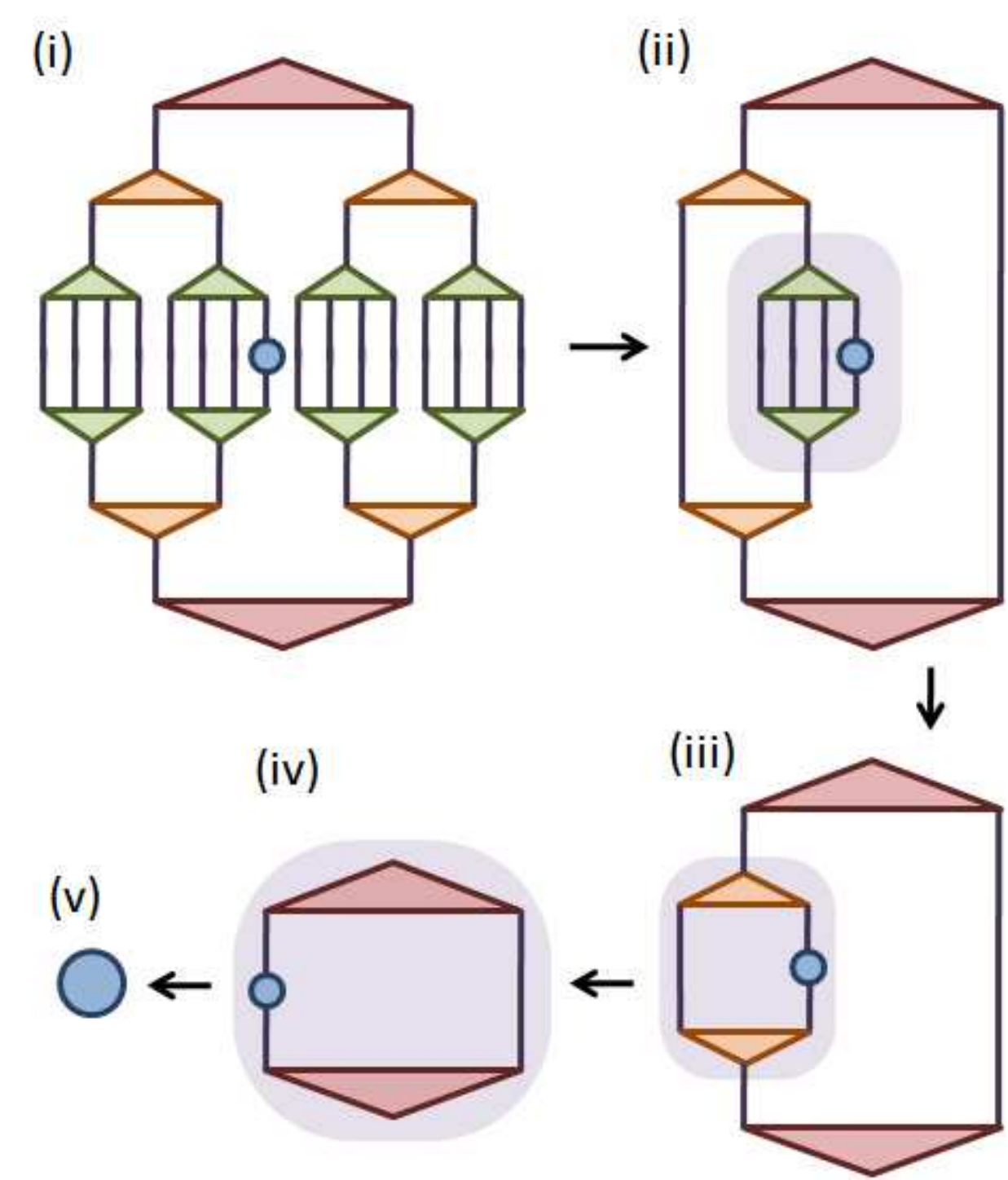}
\caption{ Computation of the expectation value $\bra{\Psi} o^{[s]}\ket{\Psi}$ of a one-site operator $o^{[s]}$ acting on site $s\in\mathcal{L}$. (i) Tensor network to be contracted. (ii) Tensor network left after many of the isometries are annihilated by their Hermitian conjugate, see Fig. \ref{fig:Isometric}. {After the steps from  (iii) to (v)  the expectation value is obtained.} } 
\label{fig:OneSiteEval}
\end{center}
\end{figure}

\begin{figure}[!tb]
\begin{center}
\includegraphics[width=6cm]{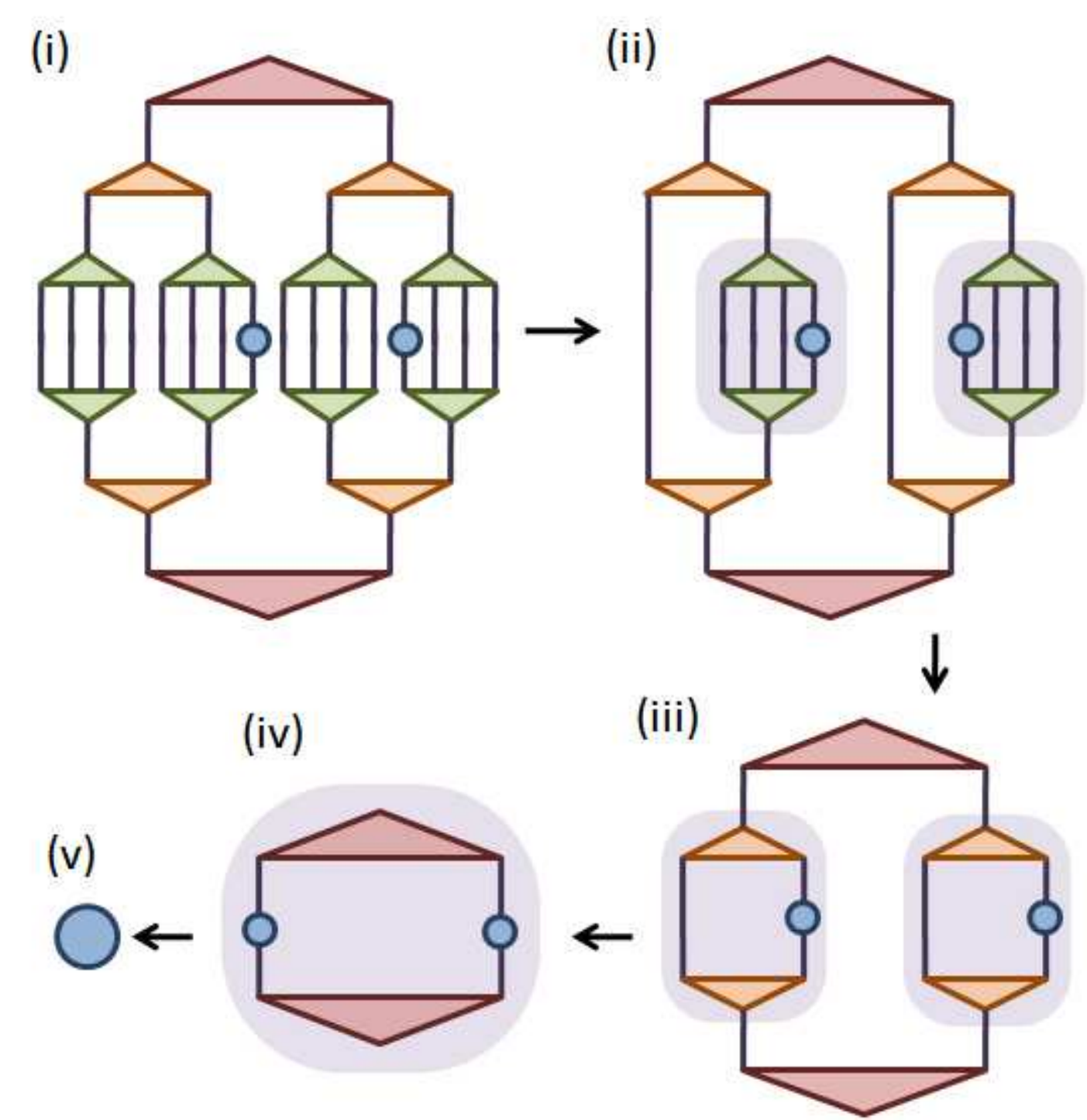}
\caption{ Computation of the expectation value $\bra{\Psi} o^{[s]}o^{[s']}\ket{\Psi}$ corresponding to a two-site correlation function. (i) Tensor network to be contracted. (ii) Tensor network left after several isometries are annihilated by their Hermitian conjugate. {After the steps from  (iii) to (v)  the expectation value is obtained.}} 
\label{fig:TwoSiteEval}
\end{center}
\end{figure}

\begin{figure}[!tb]
\begin{center}
\includegraphics[width=6cm]{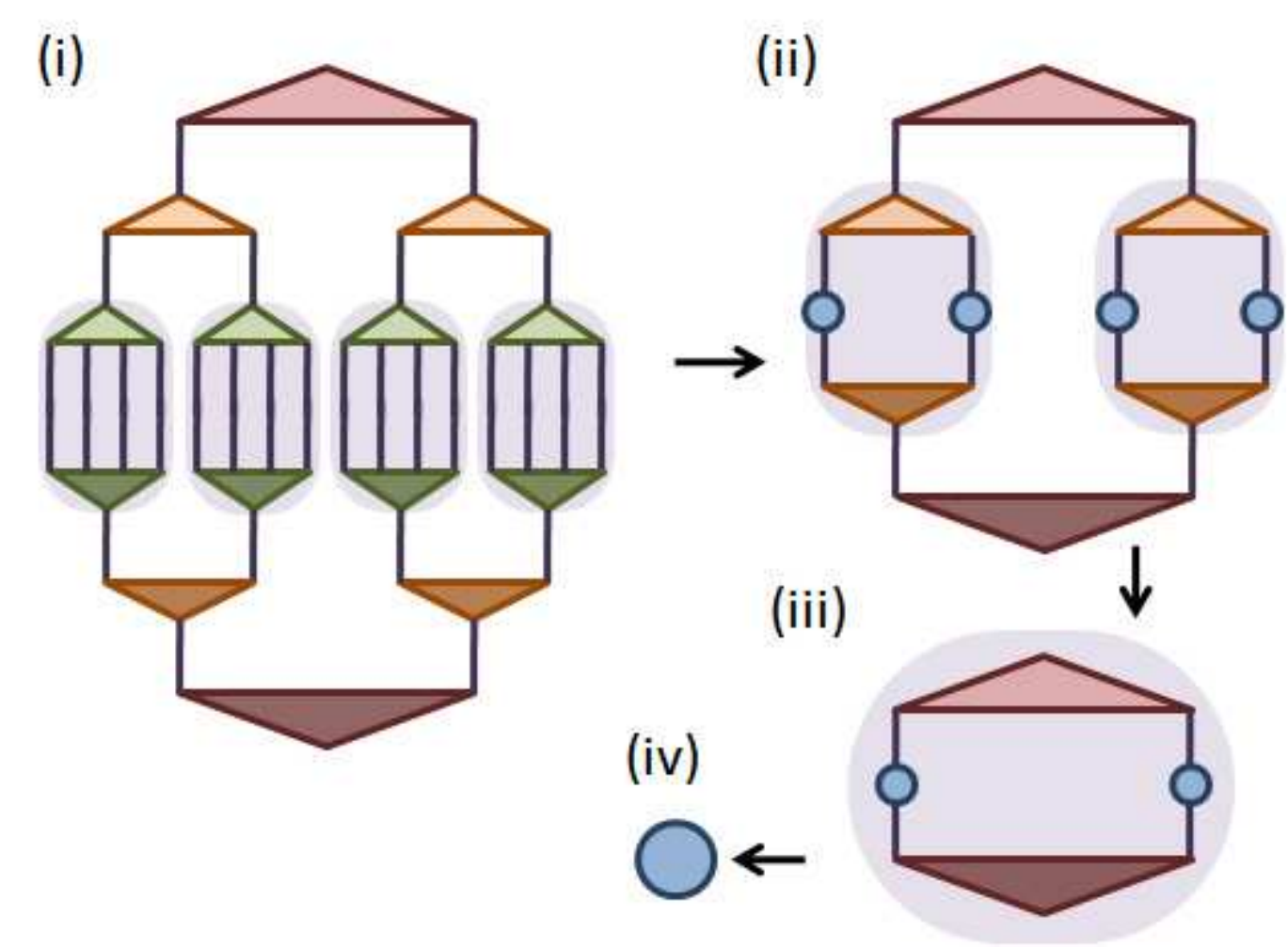}
\caption{Computation of the overlap or fidelity $\braket{\Psi_1}{\Psi_2}$ between two states $\ket{\Psi_1}$ and $\ket{\Psi_2}$ each represented with a TTN. (i) Tensor network corresponding to  $\braket{\Psi_1}{\Psi_2}$. Notice that in this case no isometry is annihilated, since the isometries of the two TTNs are not the same. {After the steps from  (ii) to (iv)  the the overlap or fidelity is obtained.} }
\label{fig:FidelityEval}
\end{center}
\end{figure}

\subsection{Relation to MERA}

The TTN ansatz considered in this work can also be thought of as a particular case of the MERA, see Refs. \cite{vidal_entanglement_2007,vidal_class_2006,evenbly_algorithms_2007,evenbly_entanglement_2008}. Specifically, a MERA where disentanglers are replaced with identity operators becomes a TTN. We emphasise that the manipulations of a TTN, as discussed in Sects. III and IV, are different than those for the MERA. Indeed, the absence of disentanglers changes the optimal pattern of tensor network contractions. As a result, for instance, the scaling of the computational cost with $\chi$ is much smaller with a TTN (namely $\mathcal{O}(\chi^4)$) than with a MERA \cite{evenbly_entanglement_2008}. Of course, for large 2D lattices the benefits of having a cost that scales as a   smaller power of  $\chi$ are offset by the need to use a much larger value of $\chi$, and simulations with a TTN become more expensive than with the MERA.

\section{Computation of local operators, fidelities and entropies}
\label{sect:computation}

In this section we assume that a TTN for the state $\ket{\Psi}$ of an $L\times L$ lattice $\mathcal{L}$ has been provided, and explain how to extract a number of quantities of interest from it. 

This section is presented before explaining the optimisation algorithm in the next section mostly for two reasons. On the one hand, the algorithm of Sect. \ref{sect:algorithm} is only one of many possible ways of obtaining a TTN (one could alternatively consider using a different optimisation algorithm \cite{friedman_density_1997,lepetit_density-matrix_2000,martn-delgado_density-matrix_2002,shi_classical_2006,nagaj_quantum_2008}, or obtain a TTN through an analytical derivation \cite{fannes_ground_1992,niggemann_quantum_1997}) and yet in all cases it is still necessary to extract  information from the TTN representation. On the other hand, by explaining now  how to compute quantities of interest from a TTN, we also introduce material that will be useful later on in order to understand the optimisation algorithm.

\subsection{Expectation value of local operators, two-point correlation functions and fidelity}

We start by noticing that since the TTN is made of isometries, the state $\ket{\Psi}$ it represents is automatically normalised, $\braket{\Psi}{\Psi} = 1$.


Given a local operator $o^{[s]}$ that acts on a single site $s$ of $\mathcal{L}$, the expectation value 
\begin{equation}
	\bra{\Psi} o^{[s]} \ket{\Psi}
\end{equation}
can be computed by contracting the tensor network of Fig. \ref{fig:OneSiteEval}. Notice that an important fraction of the isometries in the TTN are annihilated by   their Hermitian conjugate pair, see Fig. \ref{fig:Isometric}, and are therefore not involved in the computation of $\bra{\Psi} o^{[s]} \ket{\Psi}$. 

A local operator $o^{[ss']}$ that acts on two sites $s$ and $s'$ of $\mathcal{L}$ can always be decomposed as a sum of products of one-site operators $o^{[s]}_{\alpha}$ and $o^{[s']}_{\beta}$,
\begin{equation}
	o^{[s s']} = \sum_{\alpha \beta} o^{[s]}_{\alpha} o^{[s']}_{\beta}. 
\end{equation}
Therefore, without loss of generality we can concentrate on the calculation of a two-point correlation function
\begin{equation}
	\bra{\Psi} o^{[s]}o^{[s']} \ket{\Psi}
\end{equation}
This computation is achieved by contracting the tensor network of Fig. \ref{fig:TwoSiteEval}. A minor difference with the previous contraction for a single-site operator is that now less pairs of isometries are annihilated. 

More generally, the expectation value of a product of $p$ one-site operators \bra{\Psi }$o^{[s_1]}o^{[s_2]} \cdots o^{[s_p]} \ket{\Psi}$ can also be obtained by contracting a similar tensor network, and so can the overlap or fidelity $\braket{\Psi_1}{\Psi_2}$ between two states $\ket{\Psi_1}$ and $\ket{\Psi_2}$ represented by a TTN with equivalent tree structure, see Fig. \ref{fig:FidelityEval}. 

\subsection{Spectrum and entropy of the reduced density matrix of  a block of sites}

\begin{figure}[!tb]
\begin{center}
\includegraphics[width=8cm]{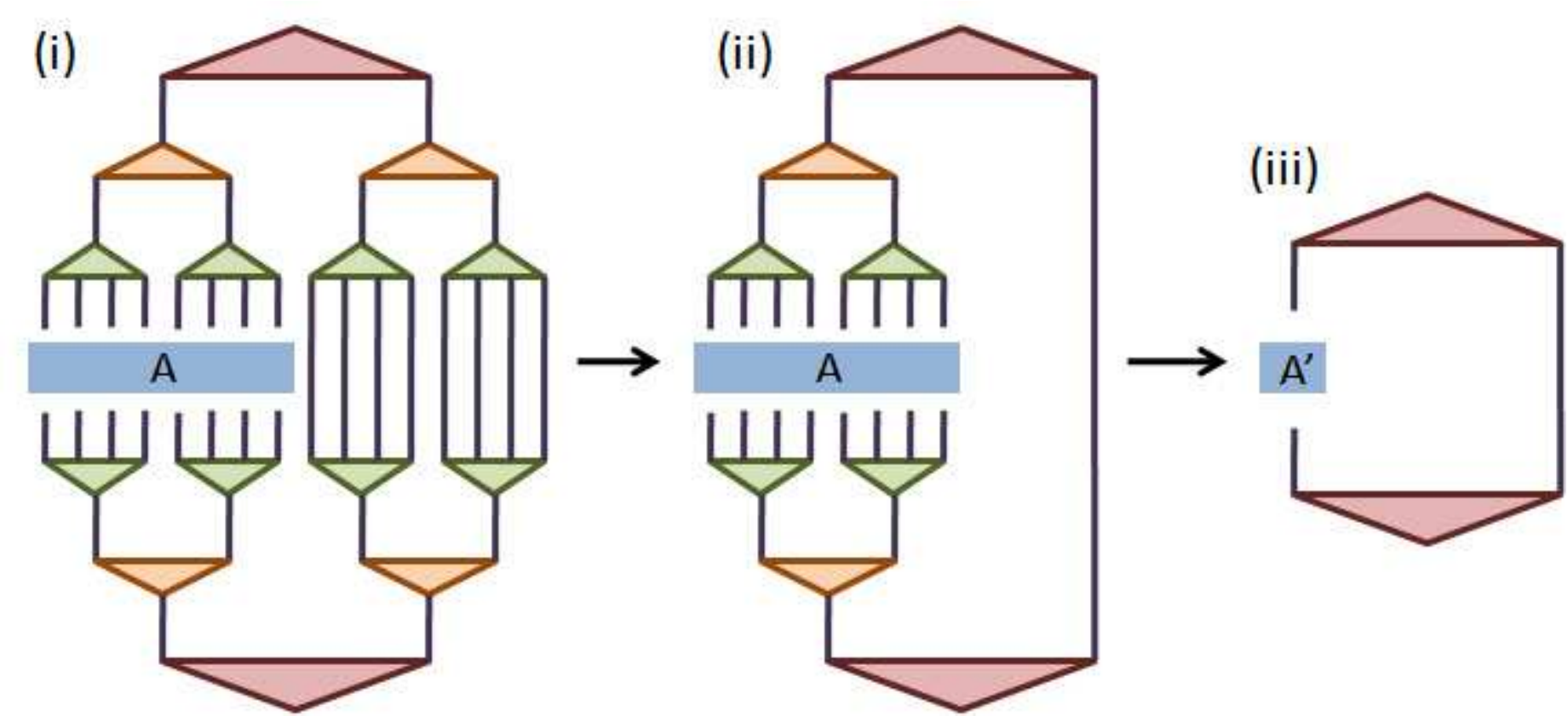}
\caption{Computation of the spectrum $\{p_{\alpha}\}$ of the reduced density matrix $\rho^A$ for a block $A$ that corresponds to one of the coarse-grained sites. (i) Tensor network corresponding to $\rho^A$ where $A$ is half of the lattice. (ii) Tensor network left after several isometries are annihilated with their Hermitian conjugate. (iii) since the spectrum of $\rho^{[A]}$ is not changed by the isometries acting on $A$, we can also eliminate those isometries and we are left with a network consisting of only two tensors, which can now be contracted together.} 
\label{fig:SpectEval}
\end{center}
\end{figure}

Finally, from the TTN it is straightforward to compute the spectrum $\{p_{\alpha}\}$ of the reduced density matrix $\rho^A$ (cf. Eq. \ref{eq:rhoA}) for certain blocks $A$ of sites, namely those that correspond to an effective site of any of the coarse-grained lattices $\mathcal{L}_1, \cdots, \mathcal{L}_{T-1}$. Fig. \ref{fig:SpectEval} illustrates the tensor network corresponding to $\rho^A$ for the case when $A$ is one half of the lattice. As before, many pairs of isometries are annihilated. In addition, the isometries contained within region $A$ can be removed since they do not affect the spectrum of $\rho^{[A]}$. From the spectrum $\{p_{\alpha}\}$, we can now obtain the entropy $S(A)$ of Eq. \ref{eq:entropy}.

In the large $\chi$ regime, where the bond dimension at the top layers of the TTN is much larger than in the lowest layers, the cost of contracting any of the tensor networks in Fig. \ref{fig:OneSiteEval}-\ref{fig:SpectEval} is dominated by matrix multiplications whose computational cost scales as $\chi^4$. Thus, this is the cost of all the tasks discussed in this section.

\section{Optimisation algorithm}
\label{sect:algorithm}

In this section we describe an algorithm to optimise the TTN ansatz so that it approximates the ground state $\ket{\Psi_{\mbox{\tiny GS}}}$ of a Hamiltonian $H$, 
\begin{equation}
H\ket{\Psi_{\mbox{\tiny GS}}} = E_{\mbox{\tiny GS}}\ket{\Psi_{\mbox{\tiny GS}}},
\end{equation}
defined on an $L\times L$ lattice $\mathcal{L}$ with torus topology. For simplicity we will assume that the Hamiltonian $H$ decomposes into two-site terms that couple only pairs of nearest neighbour sites $s,s' \in \mathcal{L}$, 
\begin{equation}
	H = \sum_{\langle s,s' \rangle} h^{[s,s']},
	\label{eq:Ham}
\end{equation}
although much more complicated Hamiltonians (e.g. with plaquette interactions or arbitrarily long-range interactions) can be also considered with only minor modifications of the algorithm.

\subsection{Cost function and optimisation strategy}

\begin{figure}[!tb]
\begin{center}
\includegraphics[width=3cm]{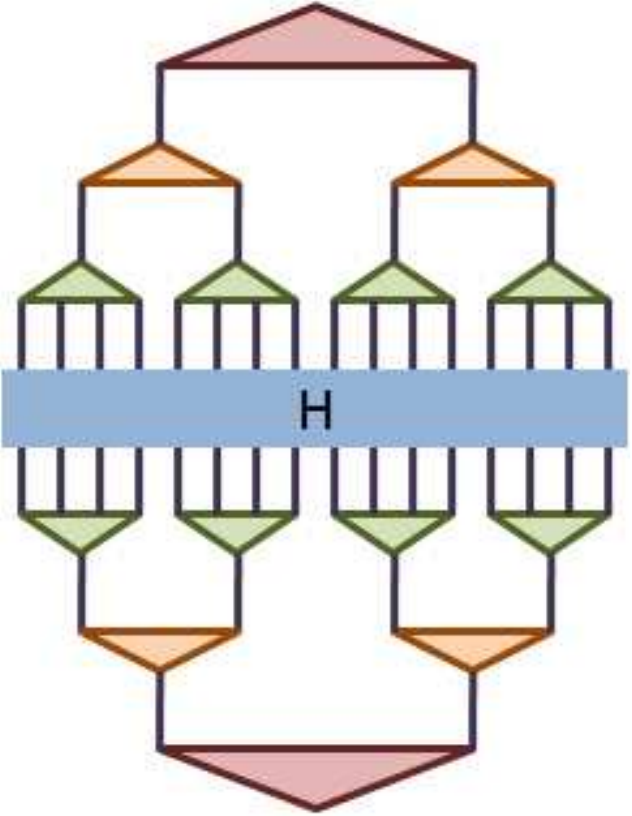}
\caption{Tensor network corresponding to the cost function $E(\{w_i\}) = \bra{\Psi_{\{w_i\}}} H \ket{\Psi_{\{w_i\}}}$ to be minimised.} 
\label{fig:FigMerit}
\end{center}
\end{figure}

Given the TTN ansatz $\ket{\Psi_{\{w_i\}}}$ at a fixed value of $\chi$, our goal is to minimise the expectation value 
\begin{equation}
	E(\{w_i\}) \equiv \bra{\Psi_{\{w_i\}}} H \ket{\Psi_{\{w_i\}}},
\end{equation}
as represented in Fig. \ref{fig:FigMerit}, by optimising all the isometries $\{w_i\}$ in the TTN, so as to obtain an optimal approximation $E(\{\bar{w}_i\})$ to the ground state energy $E_{\mbox{\tiny GS}}$,
\begin{equation}
	E(\{\bar{w}_i\}) \equiv \min_{\{w_i\} } ~ \bra{\Psi_{\{w_i\}}} H \ket{\Psi_{\{w_i\}}},
\label{eq:minimization}
\end{equation}
as well as an optimal TTN approximation $\ket{\Psi_{\{\bar{w}_i\}}}$ to the ground state $\ket{\Psi_{\mbox{\tiny GS}}}$. 

An \emph{exact} solution to Eq. \ref{eq:minimization} is not known. However, one may attempt to \emph{approximately} minimise the energy $E(\{w_i\})$ in many different ways. Here we will do so by means of an iterative optimisation strategy, which is an adaptation to the present context of the algorithm described in Ref. \cite{evenbly_algorithms_2007}. 

Starting with some  set of initially random  isometries $\{w_1, w_2, w_3, \cdots\}$, we will first optimise one of them, say $w_1$, to obtain an optimal $w_1'$. Then, given the updated set $\{w_1', w_2, w_3, \cdots\}$, we will optimise another isometry, say $w_2$, obtaining $w_2'$. In the next step, given the updated set $\{w'_1, w'_2, w_3, \cdots\}$, yet another isometry will be optimised, and so on, until we have optimised all the isometries in the TTN. This defines one sweep. Then the sweep is iterated a number of times, until the cost function $E(\{w_i\})$ is seen to converge according to some criterion, for instance until it does not change between sweeps by more than some small amount. 

\begin{figure}[!tb]
\begin{center}
\includegraphics[width=4cm]{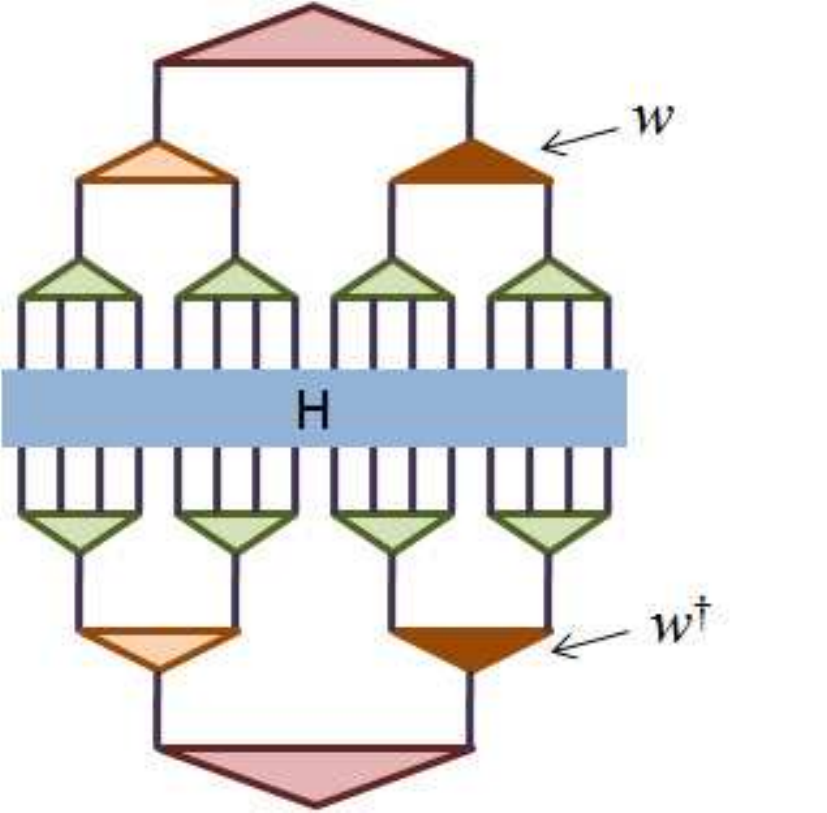}
\caption{Tensor network representation for the cost function $E(w) = F(w) + G$ in Eq. \ref{eq:Ew} depending only on one isometry $w$.} 
\label{fig:IsoEnvironment}
\end{center}
\end{figure}

\begin{figure}[!tb]
\begin{center}
\includegraphics[width=6cm]{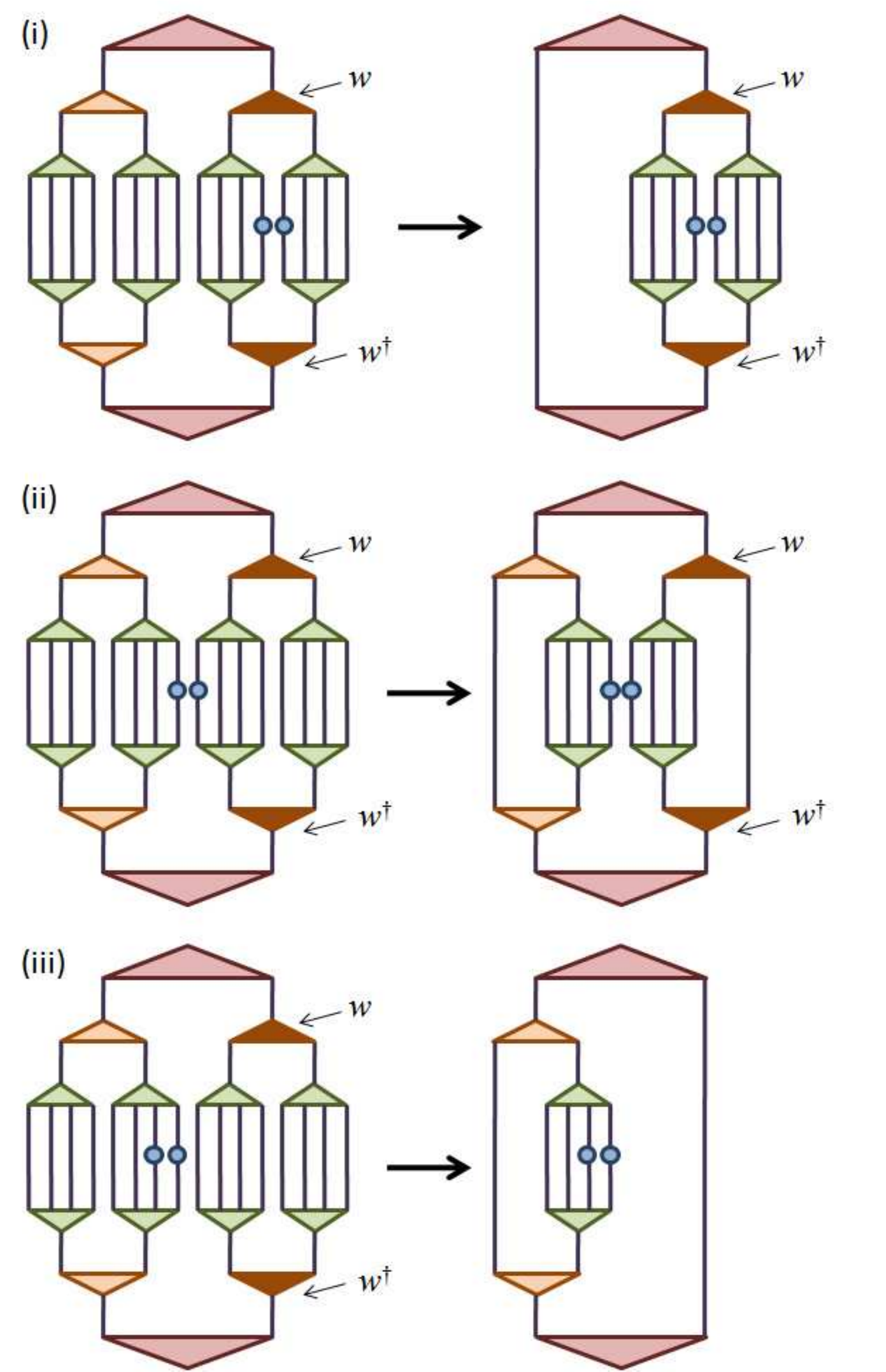}
\caption{ Examples of the three different types of two-site terms $E^{ss'}$ contributing to $E(w)$: in (i) both $s$ and $s'$ are contained within the block $A$ associated to $w$; in (ii) only one of the sites, say $s$, belongs to $A$; finally in (iii) both sites $s$ and $s'$ are outside $A$. The terms (i) and (ii) contribute to $F^{AA}(w)$ and $F^{AB}(w)$ in Eq. \ref{eq:Fw} respectively, whereas the term (iii) contributes to the constant $G$ in \ref{eq:Ew}.} 
\label{fig:IsoManyEnv}
\end{center}
\end{figure}

\begin{figure}[!tb]
\begin{center}
\includegraphics[width=6cm]{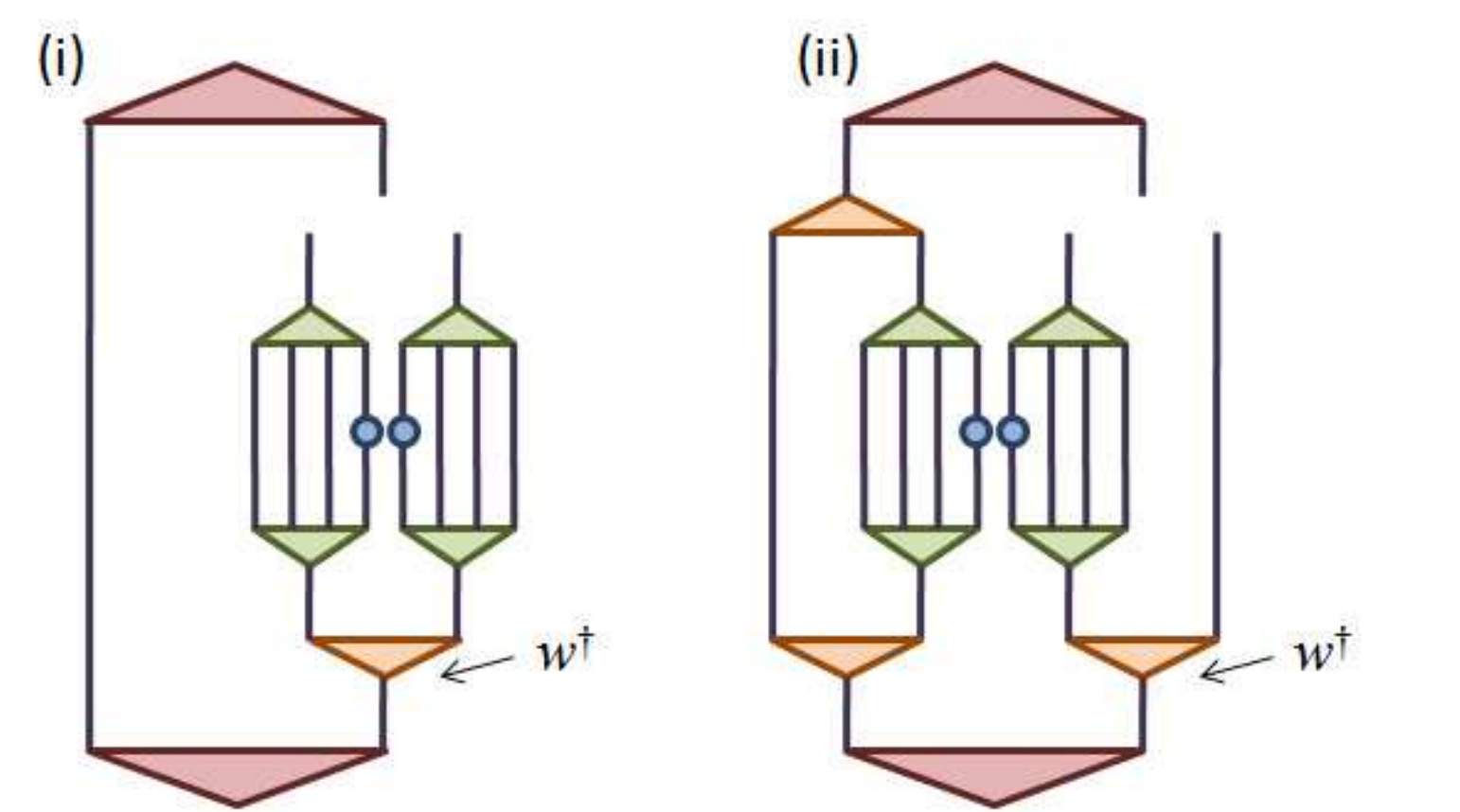}
\caption{ Examples of the two different types of two-site terms that contribute to the environment $\Upsilon$ for the isometry $w$: (i) both $s$ and $s'$ are contained within the block $A$ associated to $w$, and therefore this term contributes to $\Upsilon^{AA}$; (ii) only one of the sites, say $s$, belongs to $A$ and therefore this term contributes to $\Upsilon^{AB}$.} 
\label{fig:IsoLinearEnv}
\end{center}
\end{figure}

\subsection{Optimisation of an isometry}


Next we explain how, given a set of isometries $\{w_i\}$ for the TTN at some stage of the minimisation procedure, one can optimise one isometry $w$. Recall that $w$ is associated to a block $A$ of sites of $\mathcal{L}$.

First we notice that the cost function $E(\{w_i\})$ decomposes as a sum of two-site contributions
\begin{eqnarray}
	E(\{w_i\}) &=& \sum_{\langle s,s' \rangle} E^{ss'}(\{w_i\}) \\
	&\equiv& \sum_{\langle s,s' \rangle}\bra{\Psi_{\{w_i\}}}h^{[s,s']}\ket{\Psi_{\{w_i\}}}.
\end{eqnarray}
From now on, we also assume for simplicity in the explanation that $h^{[s,s']}$ is the product of two one-site operators. [If it is not, we can always decompose $h^{[s,s']}$ as a sum of such products.]

When viewed as a function of $w$ only, Fig. \ref{fig:IsoEnvironment}, $E(\{w_i\})$ can be divided into two pieces,
\begin{equation}
	E(w) = F(w) + G.
	\label{eq:Ew}
\end{equation}
Here $F(w)$ collects all two-site contributions $E^{ss'}$ where at least one of the two sites $s,s\in\mathcal{L}'$ are included in the block $A$ associated to $w$, and $F$ collects the rest of two-site contributions, see Fig. \ref{fig:IsoManyEnv}. Notice that if both $s$ and $s'$ in $E^{ss'}$ lie outside the block $A$, then the pair $w$ and $w^{\dagger}$ cancels out due to Eq. \ref{eq:isometry}, and $E^{ss'}$ does not depend on $w$. Therefore $G$ is independent of $w$ and we can focus on minimising $F(w)$. In turn, $F(w)$ can also be divided into two pieces,
\begin{equation}
	F(w) = F^{AA}(w) + F^{AB}(w),
	\label{eq:Fw}
	\end{equation}
where $F^{AA}(w)$ collects all contributions $E^{ss'}$ with both sites $s$ and $s'$ in $A$, whereas $F^{AB}(w)$ corresponds to the terms $E^{ss'}$ that include one site in $A$ and the other site  in its complementary $B$ (cases (i) and (ii) of Fig. \ref{fig:IsoManyEnv}). The optimisation
\begin{equation}
	\min_{w} F(w)
\end{equation}
is bilinear in $w,w^{\dagger}$ and is subject to the isometric constraint of Eq. \ref{eq:isometry}. Unfortunately, once more we do not know how to solve this minimisation exactly. 

Following Ref. \cite{evenbly_algorithms_2007}, we will approximately minimise $F(w)$ as follows. First we linearise $F(w)$ by considering $w$ to be independent of $w^\dagger$, and then we minimise the resulting cost function $I(w)= \tr(\Upsilon w)$,
\begin{equation}
	\min_{w} I(w) = \min_{w} \tr(\Upsilon w), 
\end{equation}
where $\Upsilon$ is the \emph{environment} of $w$. The function $I(w)$ can be minimised exactly, with the optimal solution corresponding to $w' = -VU^{\dagger}$, where $\Upsilon = USV^{\dagger}$ is the singular value decomposition of $\Upsilon$.

Once we have obtained the optimal $w'$, we can replace $w^{\dagger}$ with $w'^{\dagger}$ in $F(w)$, resulting in an updated environment $\Upsilon'$ that we use to minimise $I(w)$ again. Iteration produces a sequence of isometries $\{w,w',w'', \cdots\}$ that typically lead to monotonically decreasing values of the cost function, that is $F(w) \geq F(w') \geq F(w'') \geq \cdots$. One could in principle iterate the minimisation of $F(w)$ until some level of convergence has been reached. However, in practice we only use a small number of iterations (even just one) before moving to optimise another isometry of the TTN, since in actual simulations this is seen to be already enough  to perform the minimisation of Eq. \ref{eq:minimization}. {The order in which the isometries are optimised does not seem to play a relevant role in the performance of the algorithm.}

All that is left is to explain how to compute the environment $\Upsilon$ of an isometry. Again, the environment breaks into two-site contributions corresponding to the terms $E^{ss'}$ that appear in $F^{AA}(w)$ and $F^{AB}(w)$,
\begin{equation}
	\Upsilon = \Upsilon^{AA} + \Upsilon^{AB}.
\end{equation}
Fig. \ref{fig:IsoLinearEnv} shows examples of two-site contributions to $\Upsilon^{AA}$ and $\Upsilon^{AB}$.

The cost of optimising an isometry comes from the computation of the environment $\Upsilon$ and from its singular value decomposition. These costs depend on which isometry $w$ is optimised, but the cost of sweeping over all the isometries of a given layer of the TTN can be seen to scale as $O(L\chi^{4})$, since there are $O(L)$ Hamiltonian terms $h^{[ss']}$ at the boundary between two halves of the system and computing the associated contribution to an environment $\Upsilon$ has a cost $\chi^4$. [Notice that the singular value decomposition of the environments $\Upsilon$ for the two largest isometries also costs $\chi^4$]. Therefore the leading order (in $\chi$) of the cost of sweeping over the whole tree scales as $O(L\chi^4)$, and is also proportional to the number of layers in the TTN. In a translation invariant setting where all the isometries in a layer of the TTN can be chosen to be the same, this leading cost still scales as $O(L\chi^4)$ and remains proportional to the number of layers, but it has a reduced multiplicative pre-factor. In a lattice of $8 \times 8$ sites, a typical computation of the ground state (where optimisation of isometries proceeds until the expectation value of the energy does not change by more than $10^{-10}$ per sweep) requires of the order of 100-500 sweeps.

In the case of a Hamiltonian made of long range two-site interactions, where all sites interact with all sites, the number of Hamiltonian terms $h^{[ss’]}$ grows as $O(L^4)$ and, correspondingly, the overall cost of sweeping over the whole tree scales as $O(L^4\chi^4)$.


\section{Benchmark results}
\label{sect:benchmark}

In order to test the usefulness of the TTN ansatz and to benchmark the performance of the optimisation algorithm, we consider the quantum Ising model with transverse magnetic field, as given by the Hamiltonian
\begin{equation}
  H_{\tmop{Ising}} = - \sum_{< s s' >} \sigma^{[s]}_x \otimes \sigma^{[s']}_x - \lambda
  \sigma^{[s]}_z,
\label{eq:Ising}
\end{equation}
where $\sigma_x$ and $\sigma_z$ are Pauli matrices and $\lambda$ is the magnitude of the transverse magnetic field. We consider a square lattices made of $L\times L$ sites and with toroidal boundary conditions. Since each site corresponds to a spin-$1/2$ degree of freedom, its vector space dimension is $d=2$. In the thermodynamic limit, the model is known to undergo a quantum phase transition at a value  of the transverse magnetic field $\lambda_c \approx 3.044$ \cite{rieger_application_1999,blte_cluster_2002}.

We have computed TTN approximations to the ground state of $H_{\tmop{Ising}}$ in lattices of linear size $L = \{4,6,8,10,16,32\}$ and for several values of $\chi \leq 500$. Exploiting translation invariance, we have chosen, on each layer of the TTN, all isometries to be the same. For $L \leq 8$ we are in a \emph{quasi-exact regime} where results appear to be very accurate, whereas for $L \geq 10$ we are in an \emph{approximate regime} where the results are not yet converged with respect to $\chi$, but it is still possible to obtain qualitatively correct results.

\begin{figure}[!tb]
\begin{center}
\includegraphics[width=8cm]{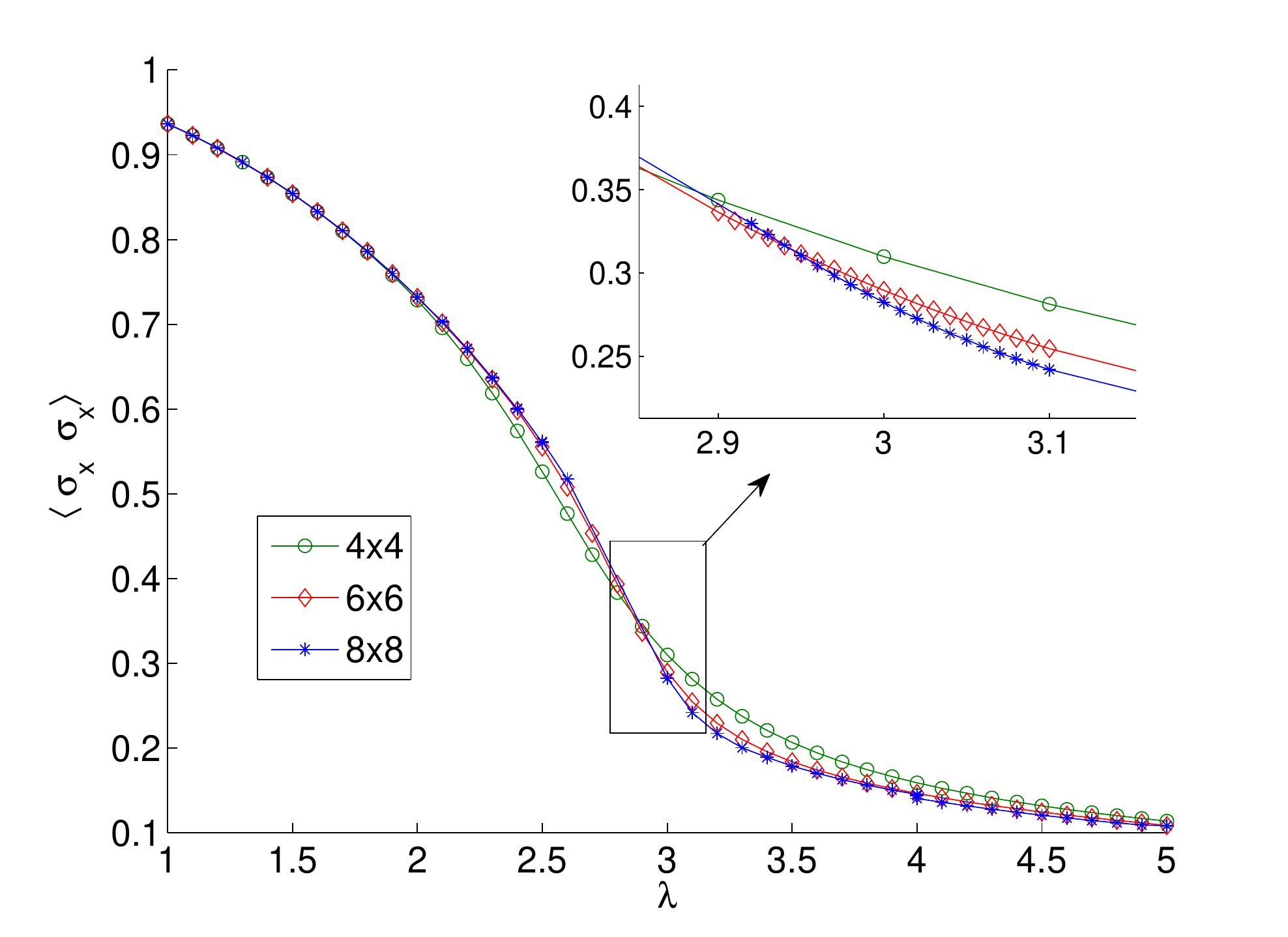}
\caption{Expectation value $\langle \sigma_x \sigma_x\rangle$ as a function of the transverse magnetic field $\lambda$ and for lattices of $4\times 4$, $6\times 6$ and $8\times 8$ sites. Notice that, as the lattice size grows, $\langle \sigma_x \sigma_x\rangle$ becomes steeper and less smooth around $\lambda \approx 3$, consistent with the existence of a critical point at $\lambda_c \approx 3.044$ in the thermodynamic limit \cite{rieger_application_1999,blte_cluster_2002} .} 
\label{fig:scanXX}
\end{center}
\end{figure}

\begin{figure}[!tb]
\begin{center}
\includegraphics[width=8cm]{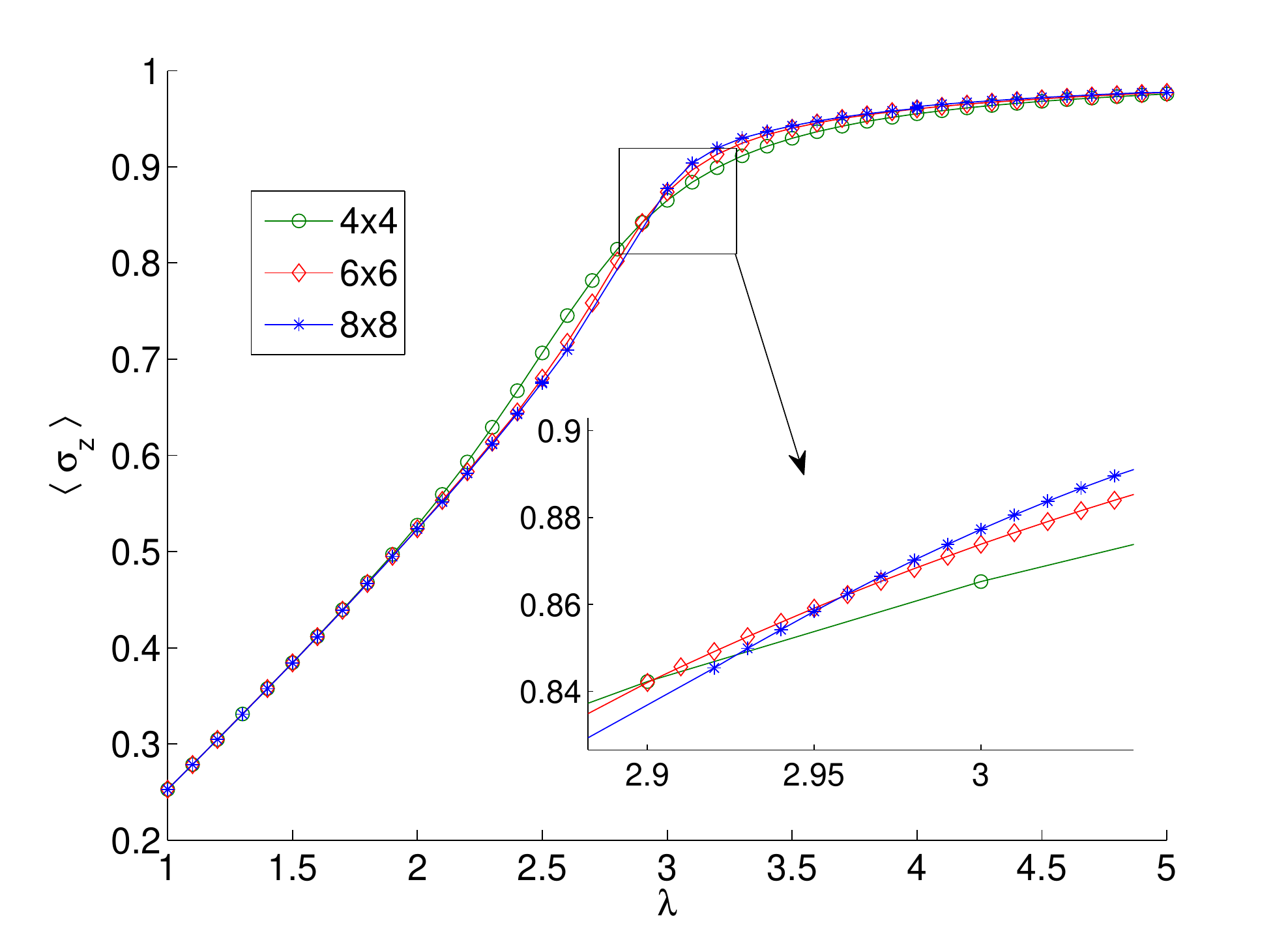}
\caption{Expectation value $\langle \sigma_{z} \rangle$ for the transverse magnetisation as a function of the transverse magnetic field $\lambda$ and for lattices of $4\times 4$, $6\times 6$ and $8\times 8$ sites. Again, as the lattice size grows $\langle \sigma_{z} \rangle $ becomes steeper and less smooth around $\lambda \approx 3$, consistent with the existence of a critical point at $\lambda_c \approx 3.044$ in the thermodynamic limit \cite{rieger_application_1999,blte_cluster_2002} .} 
\label{fig:scanZ}
\end{center}
\end{figure}

\subsection{Quasi-exact regime}

For $L={4,6,8}$ we have computed approximations to the ground state of $H_{\tmop{Ising}}$ for values of the transverse magnetic fields in the range $\lambda \in [1,5]$. Figs. \ref{fig:scanXX} and \ref{fig:scanZ} show the sxpectation values for the interaction per link
\begin{equation}
	\langle \sigma_x\sigma_x\rangle \equiv \frac{1}{2N} \sum_{\langle s,s'\rangle} \langle \sigma_x^{[s]}\sigma_x^{[s']}\rangle, 
	\label{eq:XX}
\end{equation}
and the transverse magnetisation per site
\begin{equation}
	\langle \sigma_z \rangle  \equiv \frac{1}{N} \sum_{s} \langle \sigma_z^{[s]} \rangle,
	\label{eq:Z}
\end{equation}
in terms of which the energy per site reads
\begin{equation}
	e \equiv \frac{1}{N} \langle H \rangle = -2\langle \sigma_x\sigma_x\rangle - \lambda 	\langle \sigma_z \rangle. 
	\label{eq:e}
\end{equation}
{Both observables in Figs. \ref{fig:scanXX} and \ref{fig:scanZ} around  $\lambda\simeq 3$ have a very steep dependence on $\lambda$. This behaviour is consistent with the presence of a phase transition at $\lambda=3.044$ \footnote{In order to find the precise location of the transition point one should perform a finite size scaling analysis. However, at least in its simplest version, \cite{nightingale_scaling_1975,hamer_finite-size_2000}, a finite-size scaling analysis requires requires knowledge of the energy gap between the ground state and first excited state of $H$. While this gap can in principle be computed  with a TTN, such computation is beyond the scope of the present work, restricted to ground states.}  as determined in  Refs. \cite{rieger_application_1999,blte_cluster_2002}.}

\begin{figure}[!tb]
\begin{center}
\includegraphics[width=8cm]{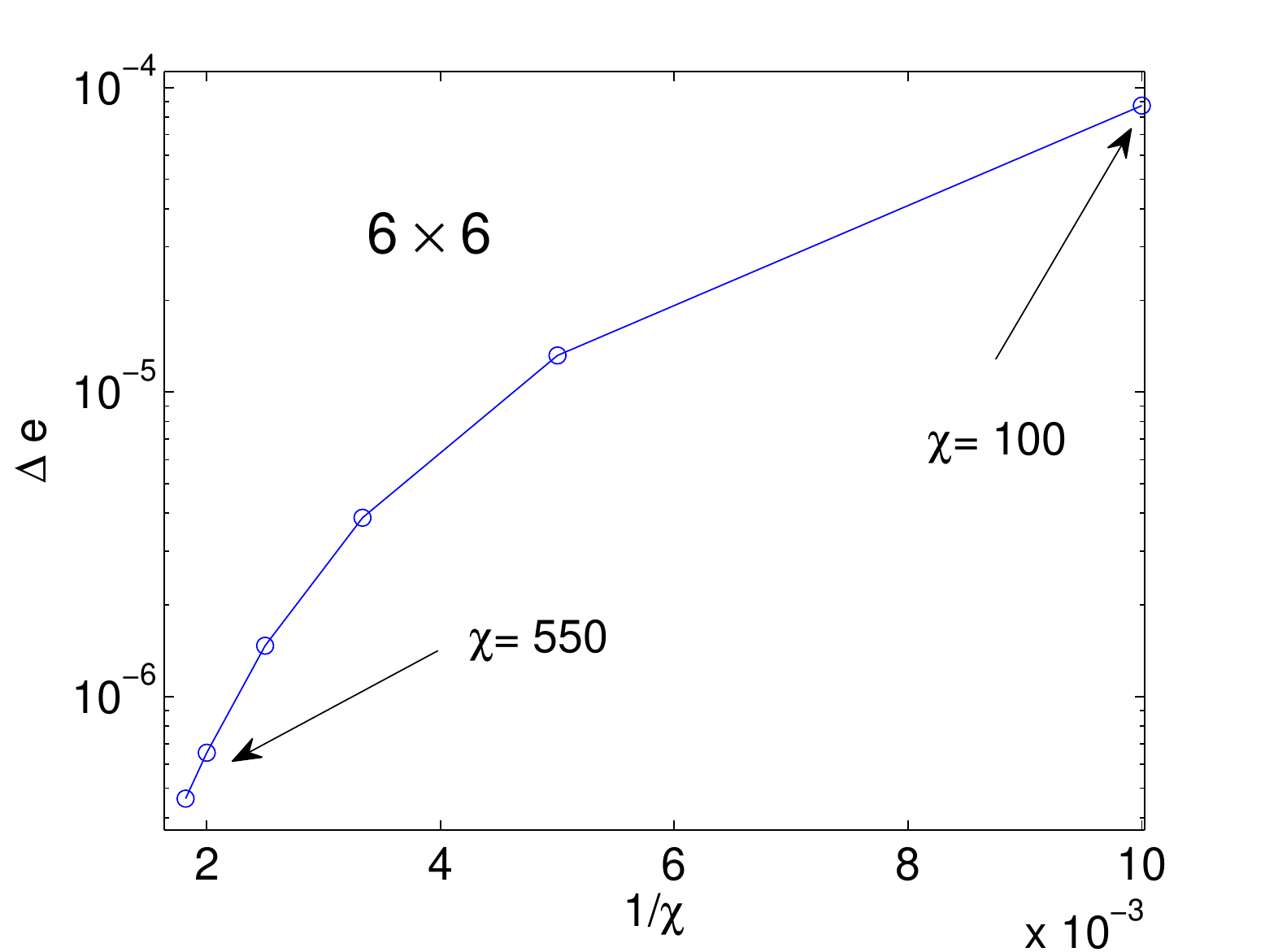}
\caption{Error of the ground state energy per site $\Delta e \equiv e(\chi) - e_{exact}$ (cf. Eq. \ref{eq:e}) for a  $6\times 6$ lattice   plotted as a function of $1/\chi$ at $\lambda = 3.05266$.  $\chi$ varies in the range from $\chi=100$ to $\chi=550$. $e_{exact}$ is extracted from  Ref. \cite{hamer_finite-size_2000}. In the plot we see that  $\Delta e$ ranges from $\Delta e \sim \mathcal{O}(10^{-4})$ in the  case of $\chi=100$ to  $\Delta e \sim \mathcal{O}(10^{-7})$ in the case of $\chi=550$. All the TTN simulations used here require much smaller computational resources than the full exact diagonalisation calculation as we explain in detail in the main text.} 
\label{fig:comp_exact}
\end{center}
\end{figure}

In order to assess the accuracy of our numerical results, we first compare the expectation value of the energy with its exact value, as obtained in Ref.\cite{hamer_finite-size_2000}   using exact diagonalisation techniques on a $6 \times 6$ lattice. [Notice that $6 \times 6$ is the largest $L \times L$ lattice that can be addressed with exact diagonalisation]. For $\lambda=3.05266$ (and thus near the critical point) the exact value of the energy per site as obtained in \cite{hamer_finite-size_2000} is $e=-3.24727439758...$. Fig. \ref{fig:comp_exact} shows the error in the energy per site obtained with a TTN with $\chi$ ranging from $100$ to $550$. This error is of the order of $10^{-4}$ for $\chi=100$ and under $3x10^{-7}$ for $\chi=550$. In the first case, the computation lasts 20 minutes on a standard PC and uses less than 0.5 Gb of RAM. In the latter case, the simulation takes around 2 days and uses about 8 Gb of RAM. By comparison, the computation by exact diagonalisation required 35 Gb of RAM \cite{hamer_finite-size_2000}.

This means that, for the model we are considering, we obtain accurate results with a fraction of the resources needed by the exact diagonalisation algorithms. 

\begin{figure}[!tb]
\begin{center}
\includegraphics[width=8cm]{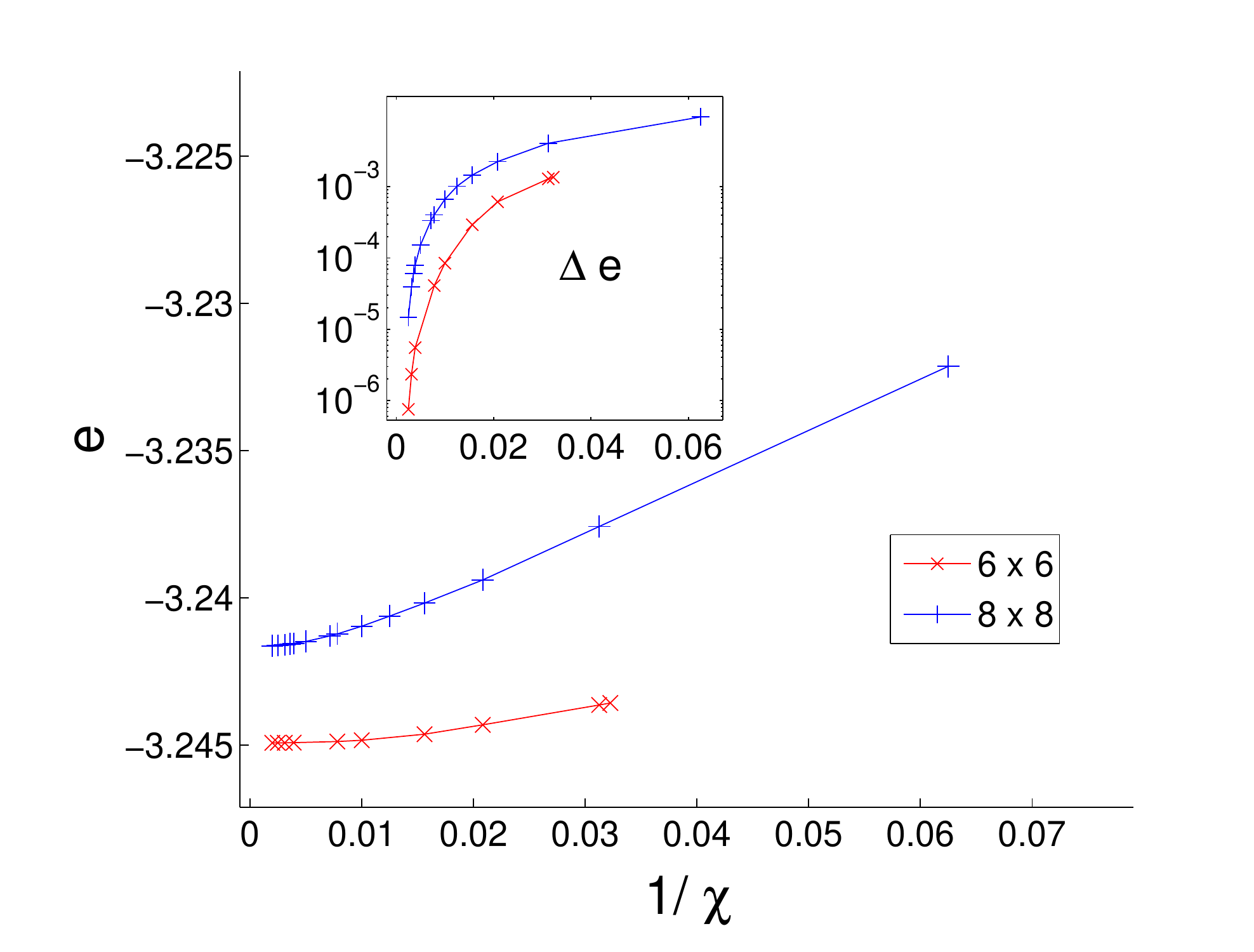}
\caption{Approximate ground state energy per site $e$ (cf. Eq. \ref{eq:e}) for $\lambda = 3.05$ plotted as a function of $1/\chi$, for lattices of $6\times 6$ and $8\times 8$ sites.  Notice that in both cases the results seem to have converged for large $\chi$ up to several digits of accuracy. The insets show the difference $\Delta e \equiv e(\chi) - e(\chi_{\max})$, as a function of $1/\chi$.} 
\label{fig:Energy}
\end{center}
\end{figure}

For larger lattices we do not have  exact results to compare against. In this case  we study the convergence in $\chi$ of the energy per site $e$ for a value $\lambda = 3.05$ of transverse magnetic field close to the critical value $\lambda_c$. This regime is the hardest to simulate, since ground states are most entangled at criticality. As shown in Fig. \ref{fig:Energy}, { where we plot the energy per site $e$ and its deviation  $\Delta e \equiv e(\chi) - e(\chi_{\max})$ } from our best estimate  $e(\chi_{\max})$, for values of $\chi$ around $500$, $e$ depends only very weakly on $\chi$. The figure also shows that, as expected, the $6\times 6$ case converges faster with large $\chi$ than the $8 \times 8$ case. 
 { Notice that, bigger systems have higher energies per site, in agreement with previous finite size scaling studies \cite{hamer_finite-size_2000} }.

\begin{figure}[!tb]
\begin{center}
\includegraphics[width=8cm]{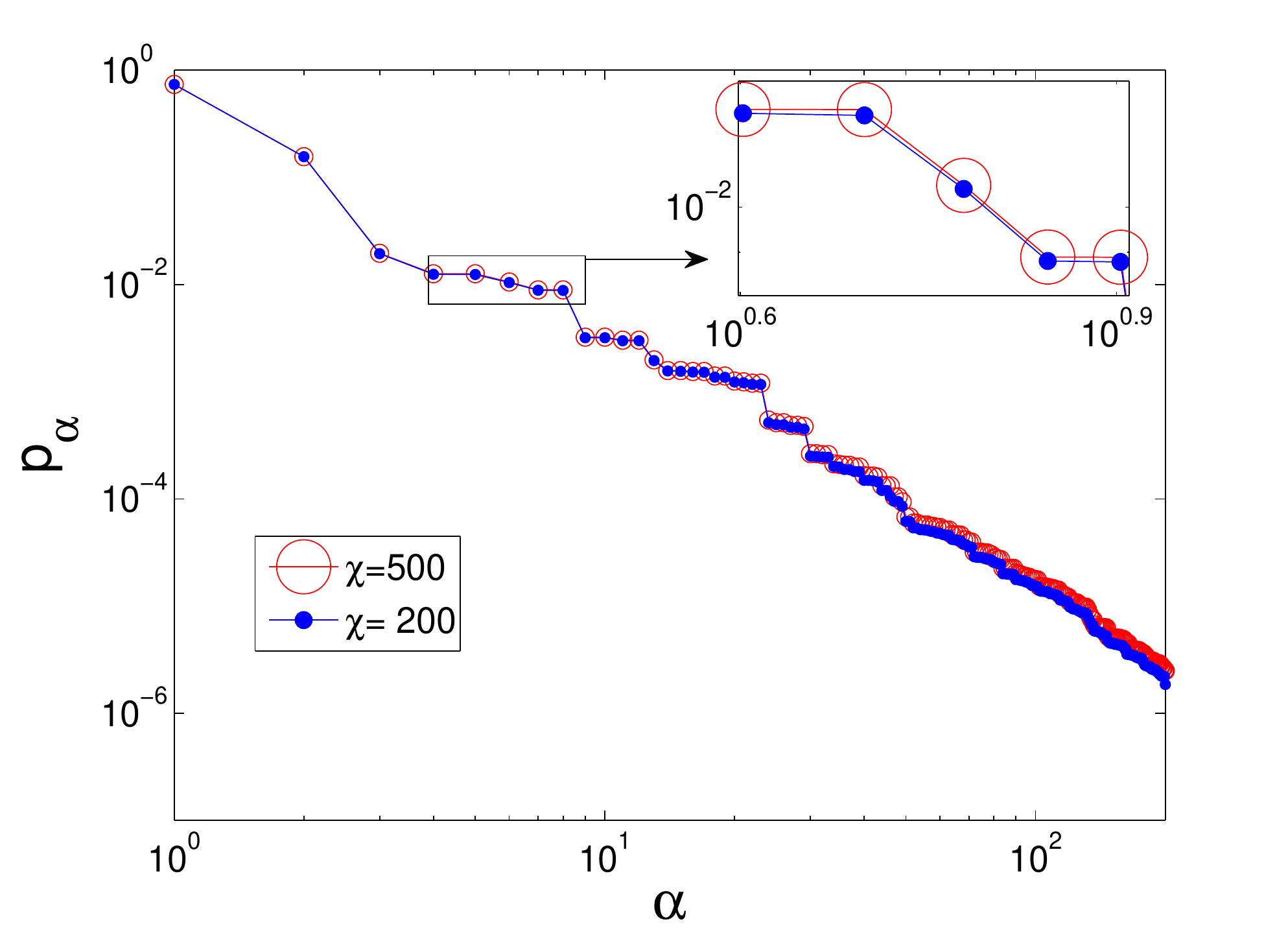}
\caption{ Spectum $\{p_{\alpha}\}$ for the reduced density matrix $\rho$ of one half of the lattice. The results the ground state of $H_{\tmop{Ising}}$ for $\lambda = 3.05$ in a $8 \times 8$ lattice. Notice the relatively fast decay of the spectrum, with e.g. $p_{\alpha} < 10^{-4}$ for $\alpha > 50$. Also, calculations with $\chi=200$ and $\chi=500$ produce spectra that are very similar for small $\alpha$. {  This is an indication that the largest eigenvalues ($p_{\alpha}$  for small $\alpha$) are already very close to their exact value.}}
\label{fig:spect8x8}
\end{center}
\end{figure}

\begin{figure}[!tb]
\begin{center}
\includegraphics[width=8cm]{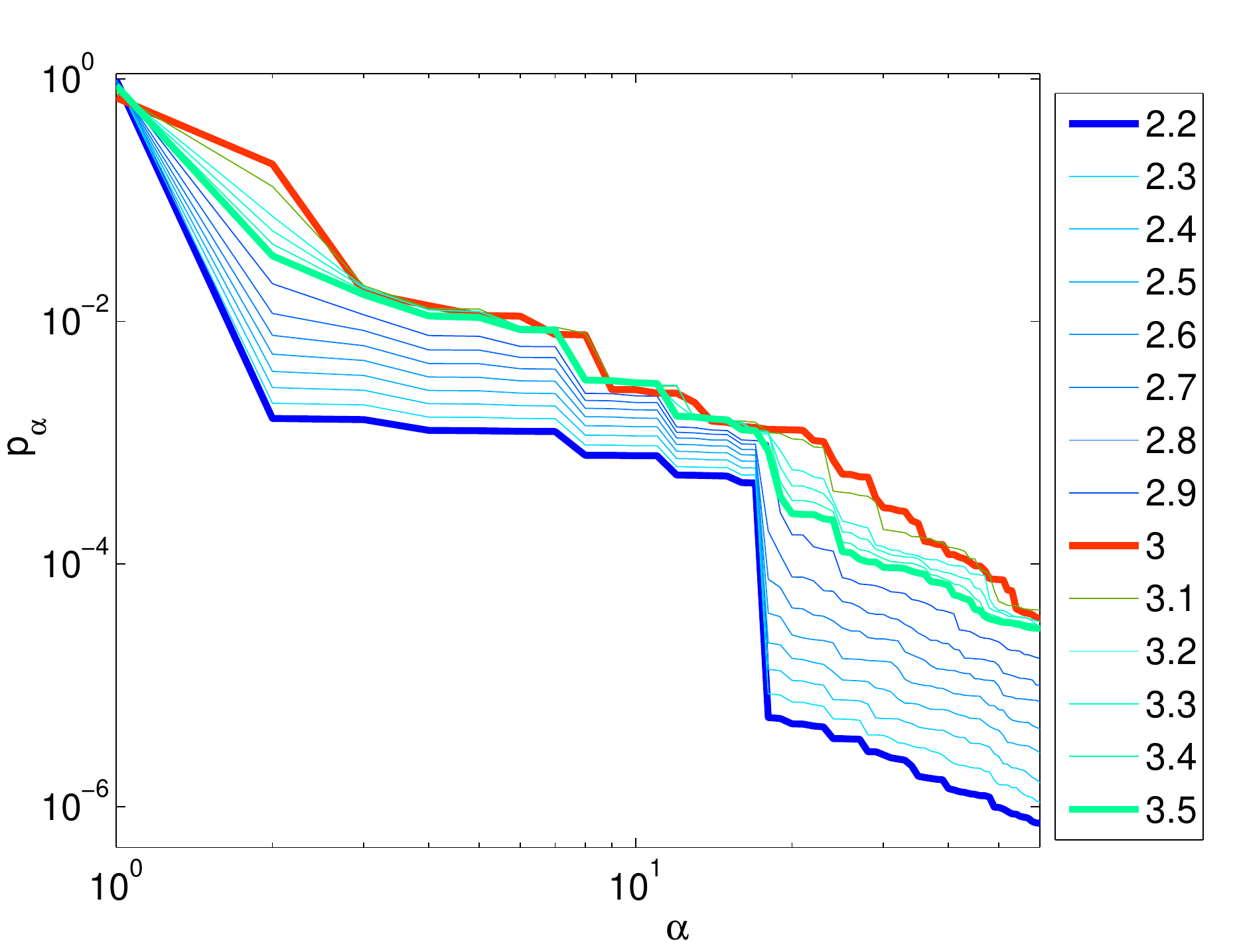}
\caption{ Spectrum of the reduced density matrix for one half the $8\times 8$ lattice for different values of $\lambda$. The calculations, conducted with $\chi = 100$, show that the spectrum decays slowest for $\lambda$ near $\lambda_c$. It also shows that for magnetic fields smaller than $\lambda_c$, the spectrum develops a clear structure of plateaux.}
\label{fig:scanSpect8x8}
\end{center}
\end{figure}

Further evidence in favour of convergence of the results in an $8\times 8$ lattice with $\chi$ is obtained by studying the spectrum of the reduced density matrix for one half of the lattice. In Fig. \ref{fig:spect8x8}  we have plotted the largest 200 eigenvalues of this spectrum, again for $\lambda = 3.05$. We see that in changing $\chi$ from $200$ to $500$ in our energy optimisation, the upper part of the spectrum remains essentially unchanged. {  Also, the spectrum $\{p_{\alpha}\}$  decays very fast as a function of $\alpha$ presenting only around 50 eigenvalues larger than $10^{-4}$. This also  implies that typical errors in the  expectation value of observables  should  be very small (we will provide examples of this statement in the section on the entropies).} The study of the spectrum of one half of the lattice as a function of $\lambda$, as displayed in Fig. \ref{fig:scanSpect8x8}, confirms that the ground state is most entangled, and therefore its computation most challenging, for $\lambda$ around $\lambda_c$. {It is also interesting to notice that, for magnetic fields $\lambda$ smaller than the critical $\lambda_c$,  the spectrum presents a very peculiar plateaux structure that  will be analysed in detail in the next section}.

\begin{figure}[!tb]
\begin{center}
\includegraphics[width=8cm]{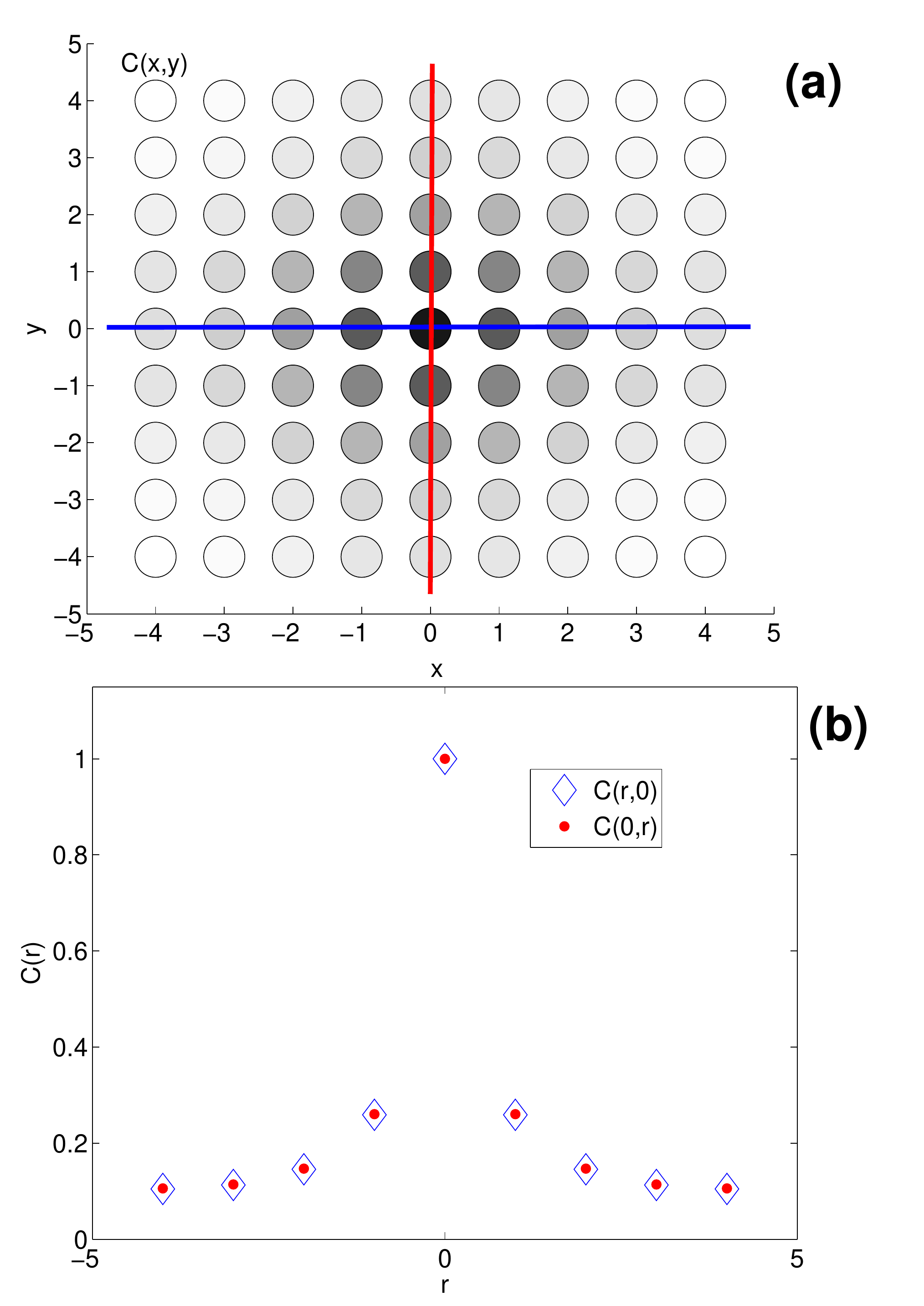}
\caption{Two-point correlation functions $C(x,y)$ of Eq. \ref{eq:correlators} for the ground state of $H_{\tmop{Ising}}$ with transverse magnetic field $\lambda = 3.05$. The correlation function between distant point in the torus remain large, as one would expect of a system that becomes critical in the thermodynamic limit.
The results, obtained with a TTN with $\chi=100$, show that the invariance of the system under $90^\circ$ rotations is preserved, in spite of the fact that the TTN manifestly breaks it at its top layers. Indeed, one can hardly distinguish $C(r,0)$ from $C(0,r)$.}
\label{fig:correlators}
\end{center}
\end{figure}

The structure of the TTN manifestly breaks translation and rotation invariance and it is natural to ask to what degree this affects the structure of correlations in the ansatz. Fig. \ref{fig:correlators} shows the two-point correlation function
\begin{equation}
	C(x,y) \equiv \langle \sigma_x^{[0,0]}\sigma_x^{(x,y)} \rangle,
\label{eq:correlators}
\end{equation}
where $(x,y)$ is a vector of integers indicating the position of a lattice site. Results obtained for a $8\times 8$ lattice with just $\chi=100$ hardly show any difference between the correlation functions in the $x$ and $y$ directions. This seems to indicate that the space symmetries expected in the ground state have already been restored at a relatively small value of $\chi$.

The results in this section demonstrate that,  for the model under
consideration,  the TTN approach offers a reliable route, based on exploiting the entropic area law, to extend the domain of quasi-exact results well beyond what is possible using exact diagonalisation techniques \cite{hamer_finite-size_2000}.

\begin{figure}[!tb]
\begin{center}
\includegraphics[width=8cm]{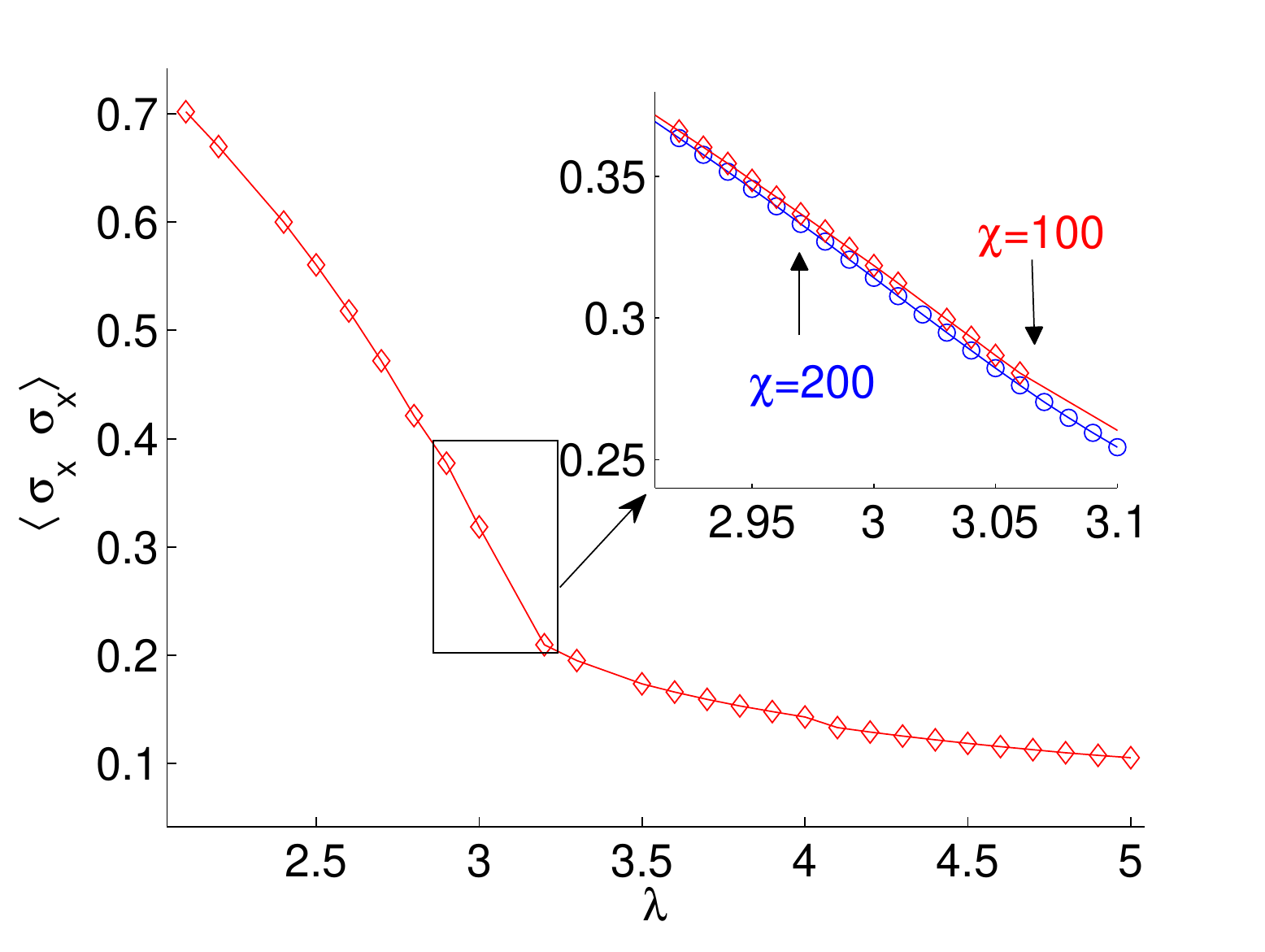}
\caption{ Expectation value $\langle \sigma_x \sigma_x\rangle$ as a function of the transverse magnetic field $\lambda$ and for a $10\times 10$ lattice. The inset shows results obtained with $\chi=100$ and $\chi=200$ for values of the transverse magnetic field $\lambda$ close to $\lambda_c$. In this approximate regime, the TTN algorithm produces results that are not converged with respect to $\chi$ near the quantum critical point.}
\label{fig:scanXX10}
\end{center}
\end{figure}

\begin{figure}[!tb]
\begin{center}
\includegraphics[width=8cm]{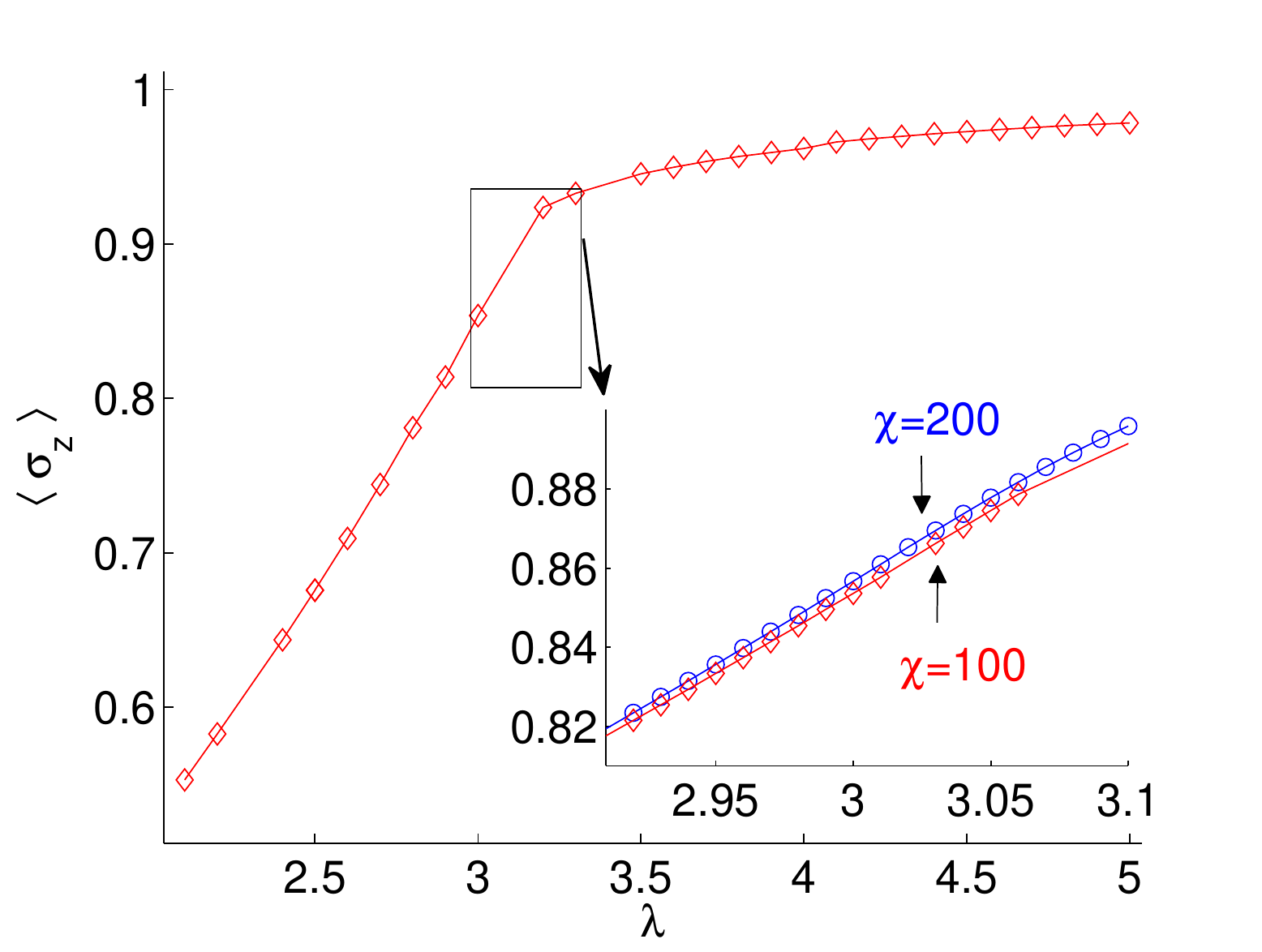}
\caption{Expectation value $\langle \sigma_z\rangle$ as a function of the transverse magnetic field $\lambda$ and for a $10\times 10$ lattice. The inset shows results obtained with $\chi=100$ and $\chi=200$ for values of the transverse magnetic field $\lambda$ close to $\lambda_c$.}
\label{fig:scanZ10}
\end{center}
\end{figure}

\begin{figure}[!tb]
\begin{center}
\includegraphics[width=8cm]{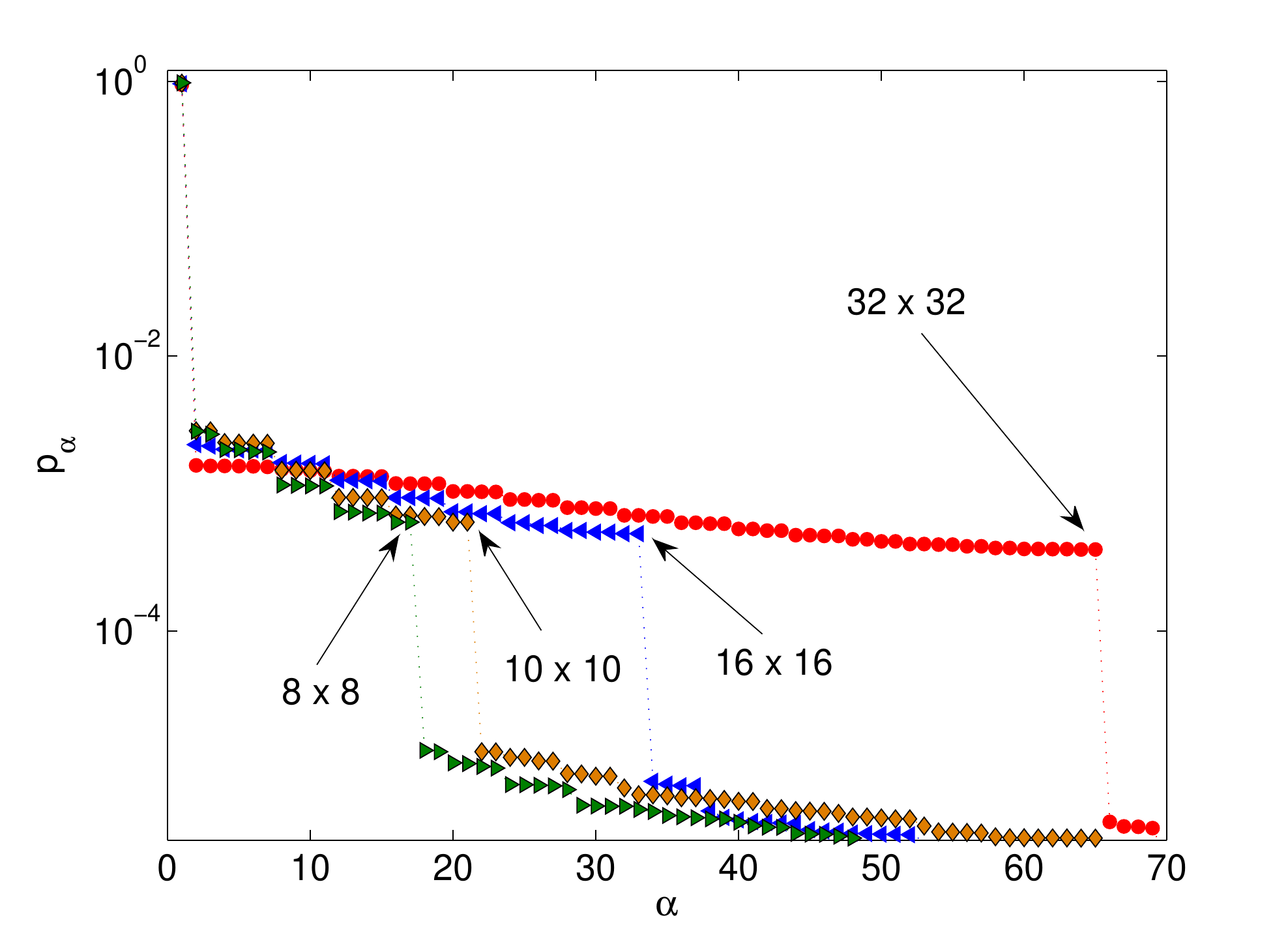} 
\caption{ Spectrum $\{p_{\alpha}\}$ of the reduced density matrix for one half of a $L\times L$ lattice in the ground state of $H_{\tmop{Ising}}$ for $\lambda=2.4$. These results, obtained with only $\chi=100$, show the presence of a plateau of exactly $2L$ eigenvalues $p_{\alpha}$, separated by two or more orders of magnitude from those of the next plateau. The structure of plateaux can be understood as a perturbative version of the entropic area law and explains why a TTN with relatively small $\chi$ can still produce converged results away from the critical point for large lattices $L\approx 10-30$. For a related discussion see also the subsection  B of section III in Ref. \cite{du_croo_de_jongh_critical_1998}.}
\label{fig:plateau}
\end{center}
\end{figure}

\begin{figure}[!tb]
\begin{center}
\includegraphics[width=8cm]{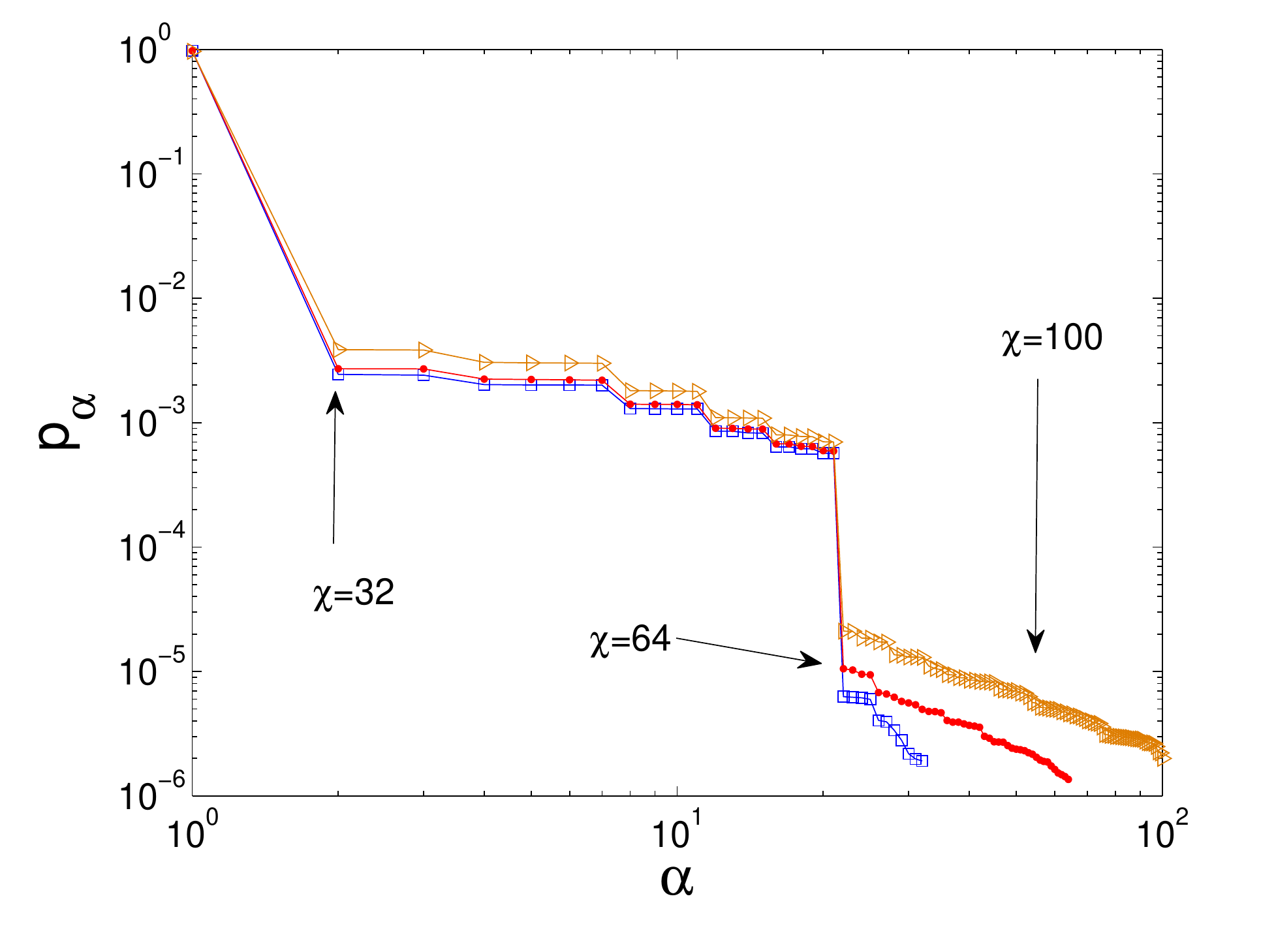}
\caption{ Spectrum $\{p_\alpha\}$ of one half of a $10\times 10$ lattice in the ground state of $H_{\tmop{Ising}}$ for $\lambda=2.4$. Results obtained with $\chi=32$, $64$ and $100$ do not differ significantly in the first $21$ eigenvalues. This shows that the presence and composition of the first plateau of $2L$ eigenvalues of Fig. \ref{fig:plateau} (with $L=10$ in this case) is robust with respect to $\chi$.}
\label{fig:stabilityPlat}
\end{center}
\end{figure}

\subsection{Approximate regime}

For lattices of linear size $L\geq 10$ we no longer obtain convincingly converged results for $\chi \approx 500$ when trying to approximate the ground state of $H_{\tmop{Ising}}$ for $\lambda$ close to $\lambda_c$. Interestingly, however, we still obtain reasonably converged results for a large range of $\lambda$ away from $\lambda_c$, which in the case of a $10 \times 10$ lattice allows us to obtain qualitatively the whole phase diagram of the system, see Figs. \ref{fig:scanXX10} and \ref{fig:scanZ10}. 

More generally, we find that converged results for lattices as large as $L=16$ and $L=32$ can be obtained, with $\chi \leq 500$, for values of $\lambda$ not too distant from $\lambda_c$. This can be explained by the presence of a plateaux structure in the spectrum of the reduced density matrix of one half of the lattice, see Fig. \ref{fig:plateau}. The first plateau consists of exactly $2L$ eigenvalues $p_{\alpha}$, that is $\alpha \in [2,2L+1]$. The second plateau is much larger, but its eigenvalues are often already very small. For instance, for $\lambda = 2.4$, the first plateau corresponds to $p_{\alpha} \approx 10^{-3}$, whereas in the second plateau  to $p_{\alpha} \approx 10^{-5}-10^{-6}$. Importantly, Fig. \ref{fig:stabilityPlat} shows that simulations with a value of $\chi$ slightly above $2L$ can already accurately reproduce the first plateau and obtain a reasonable approximation to the ground state of the system. This can be explained using perturbation theory as done e.g. in Sect. III.B in Ref. \cite{du_croo_de_jongh_critical_1998}.

\subsection{The exponential cost}
We have seen that the presence of plateaux in the spectra of reduced density matrices  can reduce the cost for an approximate description of ground states with a TTN to a linear function in the size of the system. In general, however, such short  plateaux are not expected (see e.g. the valence bond crystal example given in section \ref{subs:area_law}). In particular, close to the critical point, we do not observe any plateaux. We have repeatedly stated that a faithful representation of the ground state  with a TTN in this regime requires an exponential cost in the size of the system. Here we make this statement more precise. In order to achieve this we study how the rank of the TTN $\chi$ should increase to keep the error in the energy (as an example of a local observable) constant as we increase the system size.  The error in the energy is estimated from our numerical data as
\begin{equation}
 \Delta e (\chi) \equiv  e_{\chi}-e_{\chi_{max}}, \label{eq:epsilon}
 \end{equation} 
where $e_{\chi_{max}}$ is our best available result.

\begin{figure}[!tb]
\begin{center}
\includegraphics[width=8cm]{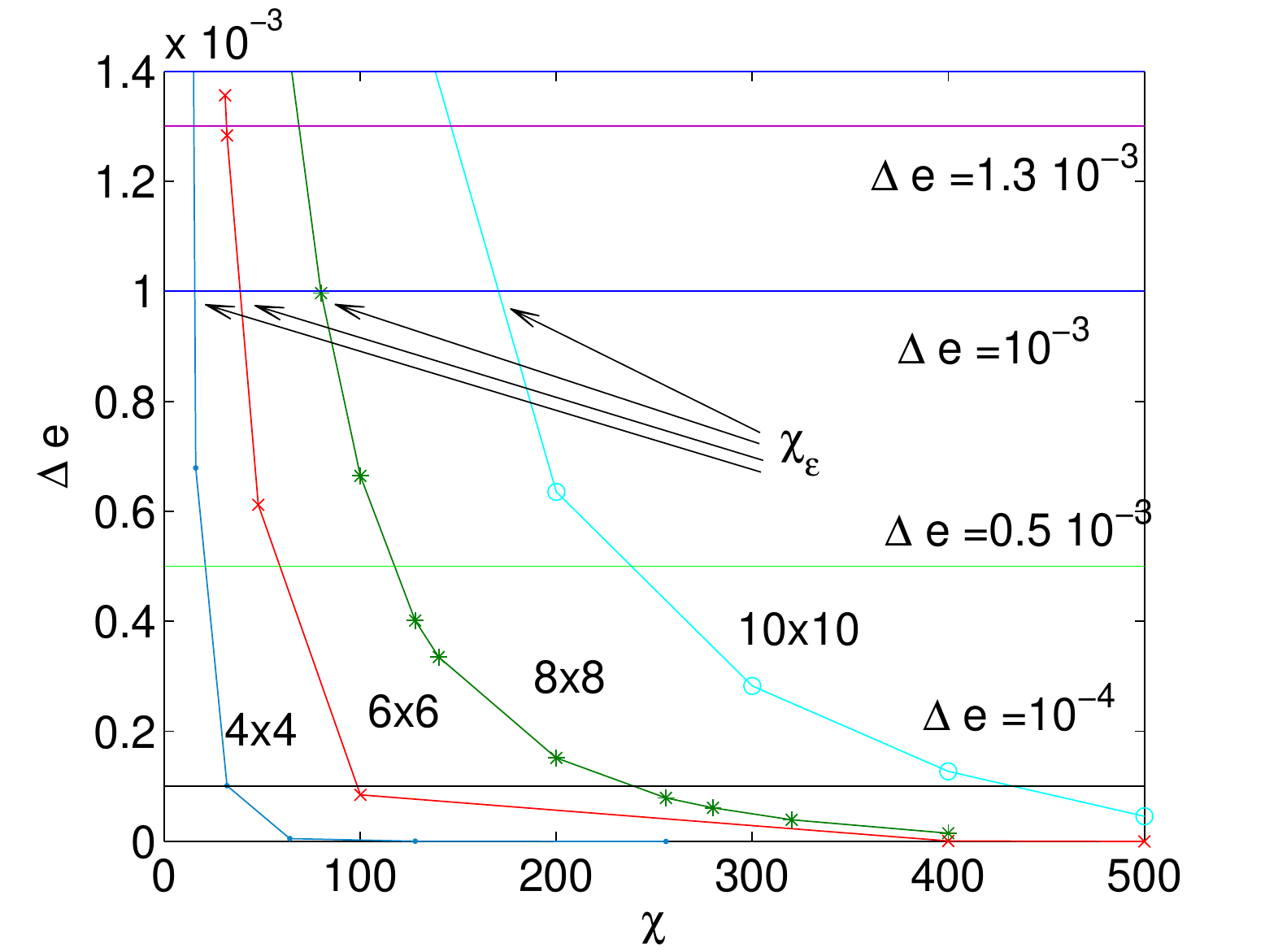}
\caption{ Plot of  $\Delta e$  defined in equation \ref{eq:epsilon} for torus of sizes $L= \{4,6,8,10\}$ and various  $\chi$.  We present several choices  of $\varepsilon$  in the range $10^{-4}\le \varepsilon \le 1.3 \  10^{-3}$ represented by horizontal lines of different colours. The arrows identify the  $\chi_{\varepsilon}$  defined in the text for the particular choice of $ \varepsilon = 10^{-3}$.}
\label{fig:intersection}
\end{center}
\end{figure}

\begin{figure}[!tb]
\begin{center}
\includegraphics[width=8cm]{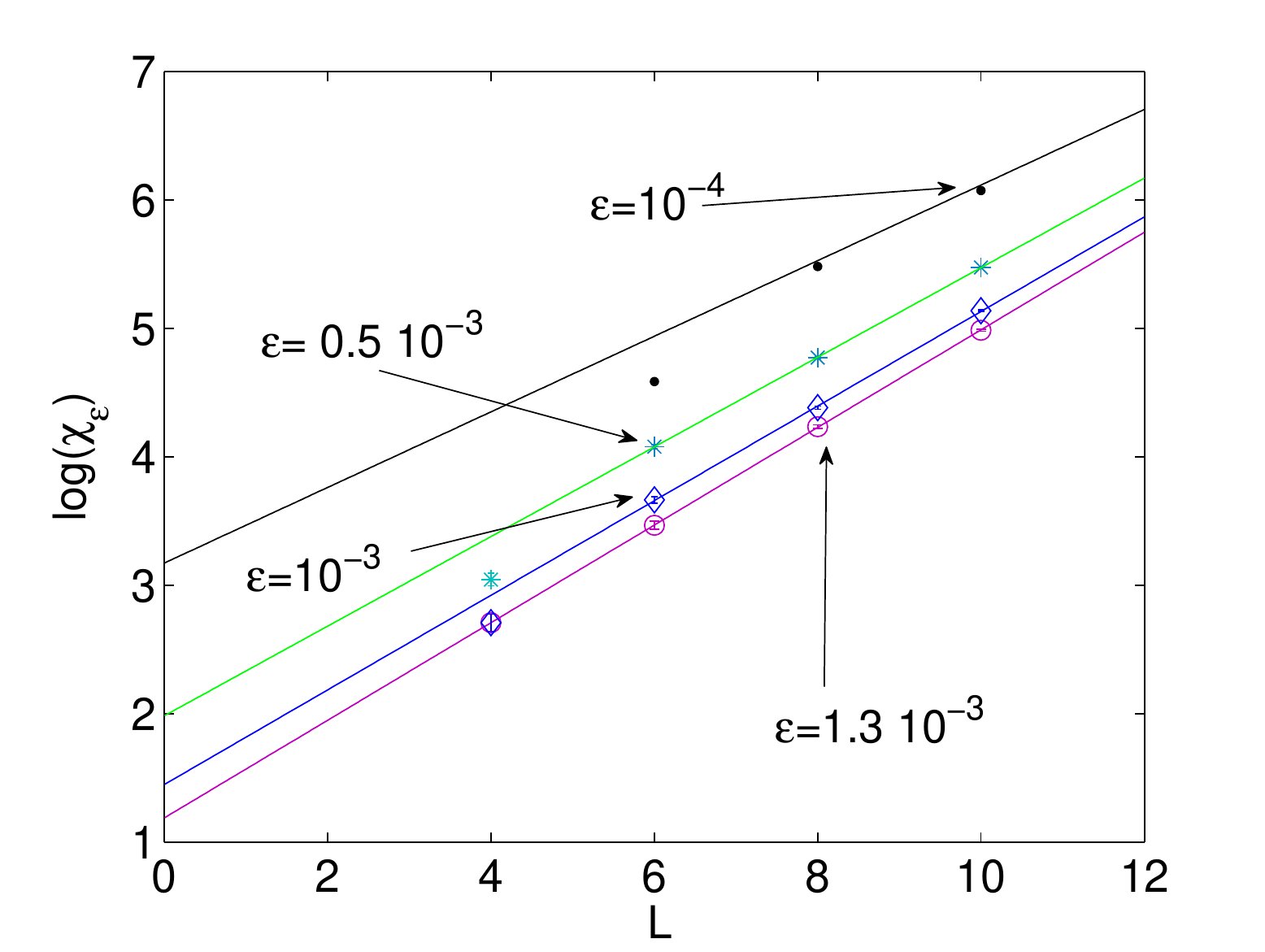}
\caption{ Plot of the logarithm of $\chi_{\varepsilon}$ as a function of $L$ for several choices of $\varepsilon$ in the range $10^{-4}\le \varepsilon \le 1.3 \  10^{-3}$.   In each case, for $L$ large enough, the  data lie on a straight line confirming that the cost of the simulation increases exponentially  with the size $L$ of the system.}
\label{fig:chiepsilon}
\end{center}
\end{figure}

For each system size $L$ we denote by  $\chi_{\varepsilon}(L)$  the minimum $\chi$  that leads to at most an error $\varepsilon$ in the energy. This procedure is illustrated in Fig.  \ref{fig:intersection}. In turn, Fig. \ref{fig:chiepsilon} displays the value of $\chi_{\epsilon}(L)$ as a function of $L$. It shows that $\chi_{\epsilon}(L)$ grows exponentially with $L$ for large $L$, thereby confirming that the cost of faithfully representing the ground state with a TTN increases exponentially with the linear size of the system.

\subsection{Error analysis}
\label{sec:error}
\begin{figure}[!tb]
 \begin{center}
   \includegraphics[width=8cm]{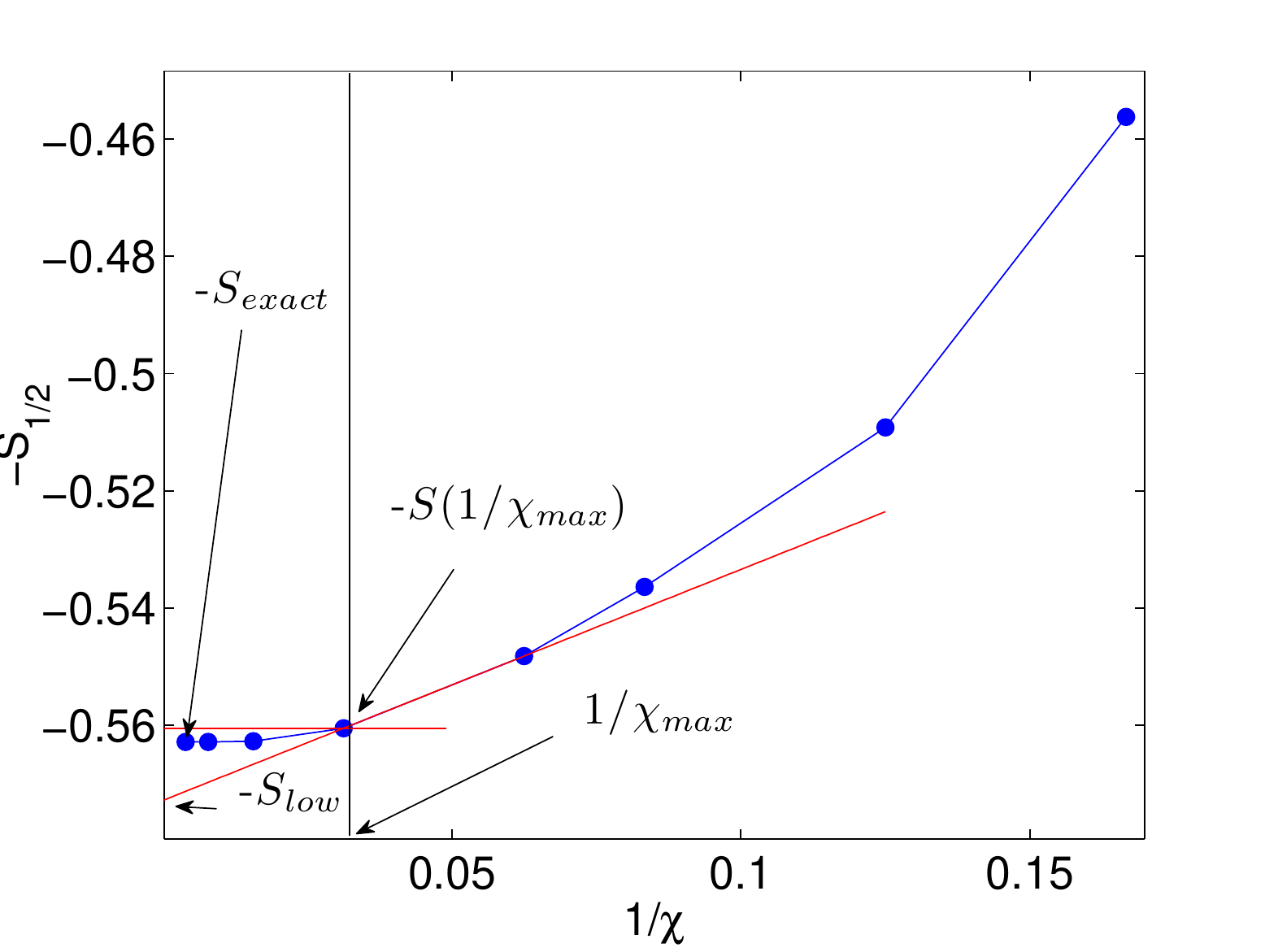} 
 \includegraphics[width=8cm]{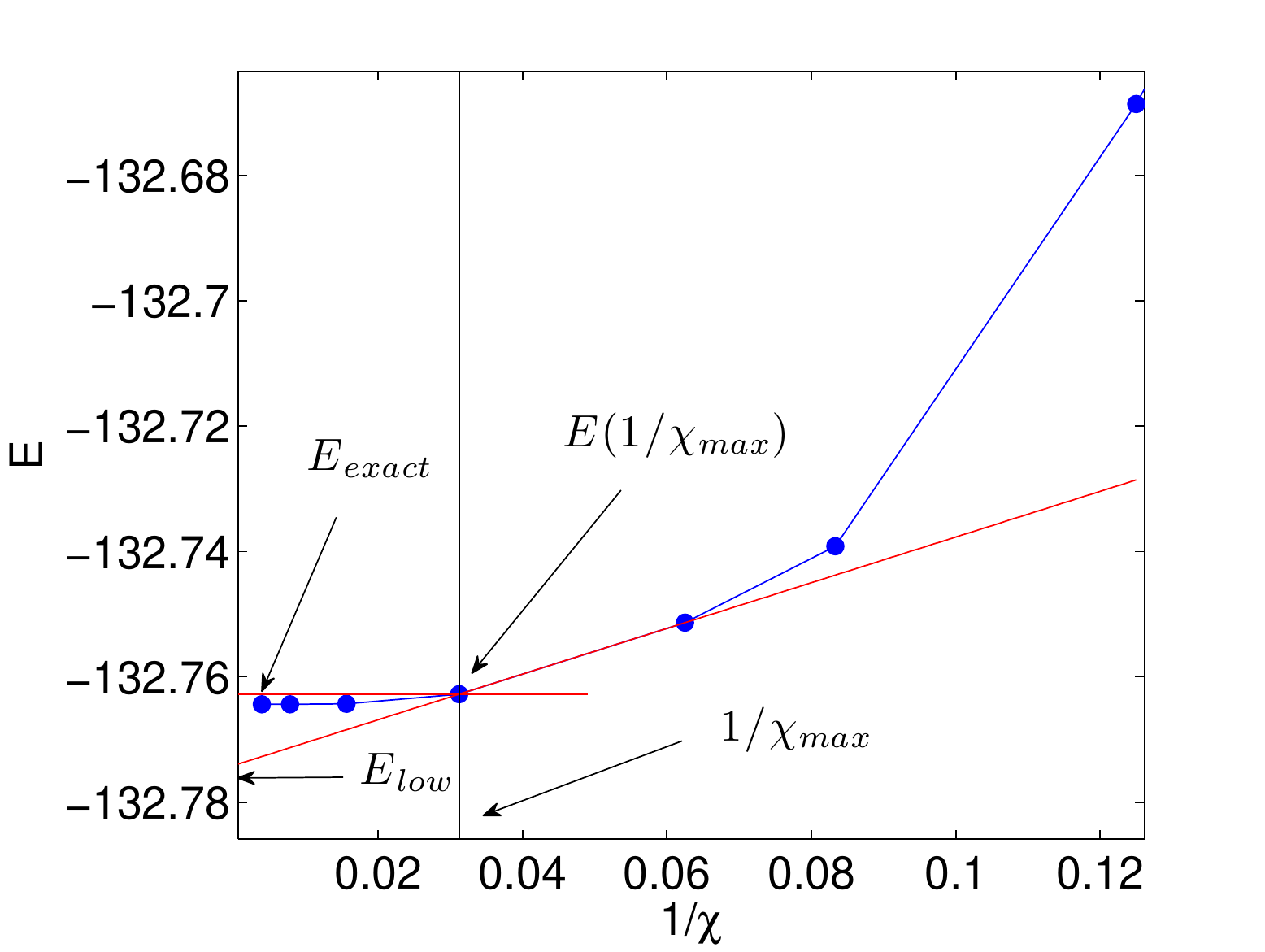}
   \caption{Analysis of  $-S_{1/2}$ (top plot) , minus  the entanglement entropy of half a torus, for the critical Ising model on a $4 \times 4$  lattice,  as a function of $1/\chi$ and the ground state energy (bottom plot). The exact result requires $\chi=256$. If we are only able to compute both observables for  $\chi\le 32$ (the region in the plots on the right of the vertical black line) the values of both observables  at $\chi=32$ provide an upper  bound to their exact results.  A lower bound is obtained  by linearly extrapolating the two observables  from the  two best values at $\chi=16$ and $\chi=32$ to $1/\chi=0$.  This is represented in the plots by the lower red line on the plots where the extrapolated values are  denoted by ${S}_{low}$ and ${E}_{low}$. In this way it is clear  that  $ -{S}_{low} \le  -S_{exact} \le  -S(1/\chi_{max}) $  and $ {E}_{low} \le  E_{exact} \le  E(1/\chi_{max}) $   as expressed in Eq. \ref{eq:extrI} in the main text. The differences $|S(1/\chi_{max})-{S}_{low}|$  and $|E(1/\chi_{max})-{E}_{low}|$ thus provide  upper bounds to the error for both $E$ and $S$ induced by considering a smaller  $\chi$ than the one required by the exact solution.
  \label{fig:increasing}}
 \end{center}
 \end{figure}

The TTN is a variational ansatz whose precision  can be  improved by increasing the value of the refinement  parameter $\chi$. Ideally, for  very large $\chi$, the results should become exact.
A precise theory on how to detect this asymptotic regime is beyond the scope of this work. However  we see  that several observables  $\mathcal{O}=  \bra{\Psi_{\chi}} O \ket{\Psi_{\chi}}$ converge to their large $\chi$ value in a characteristic way. Namely,  $\mathcal{O}(1/\chi)$ is a monotonic function of $1/\chi$  with positive, monotonically increasing derivative.
For such observables, a rough estimate of the error induced by using a finite value of $\chi$ can be obtained as follows. The monotonic nature of $\mathcal{O}$ ensures that  $\mathcal{O}(1/\chi_{max}) > \mathcal{O}(0) \equiv  \mathcal{O}_{exact} $. With a linear fit to  the behaviour of $\mathcal{O}(1/\chi)$ close to  $1/\chi_{max}$ we can extrapolate $\mathcal{O}$ to  $1/\chi=0$ and  obtain  ${\mathcal{O}}_{low} $. Indeed,  the conjectured properties of the derivative of $\mathcal{O}$ ensure that ${\mathcal{O}}_{low}  \le \mathcal{O}_{exact}$. In this way we manage to bound the exact solution  with data available from the numerical simulations
 \begin{equation}
 \label{eq:extrI}
 {\mathcal{O}}_{low} \le \mathcal{O}_{exact} \le  \mathcal{O}(1/\chi_{\max}).
 \end{equation}
As  important examples  we consider  in Fig. \ref{fig:increasing} the behaviour of both the ground state energy  and  (minus) the  entanglement entropy of the reduced density matrix of half the torus, for the critical Ising Model on a $4 \times 4$ lattice. There, we can rely on exact diagonalisation results and check that $|\mathcal{O}(1/\chi_{\max}) -{\mathcal{O}}_{low}|$ is an upper bound to the error induced by considering smaller $\chi$ than the one required by the exact solution.
In this way  we have estimated the errors appearing in the following section.
%
%

\section{Application: entropic area law}
\label{sect:entropy}
The study of the entanglement entropy for the ground states of  2D quantum systems has been the subject of several recent works  \cite{casini_universal_2007,fradkin_entanglement_2006,hsu_universal_2008,stphan_shannon_2009,fradkin_scaling_2009,yu_entanglement_2008,casini_entanglement_2009,metlitski_entanglement_2009}. 
The TTN approach provides a natural scenario for these  studies.

\subsection{Entanglement entropy}
\begin{figure}[!tb]
\begin{center}
\includegraphics[width=8cm]{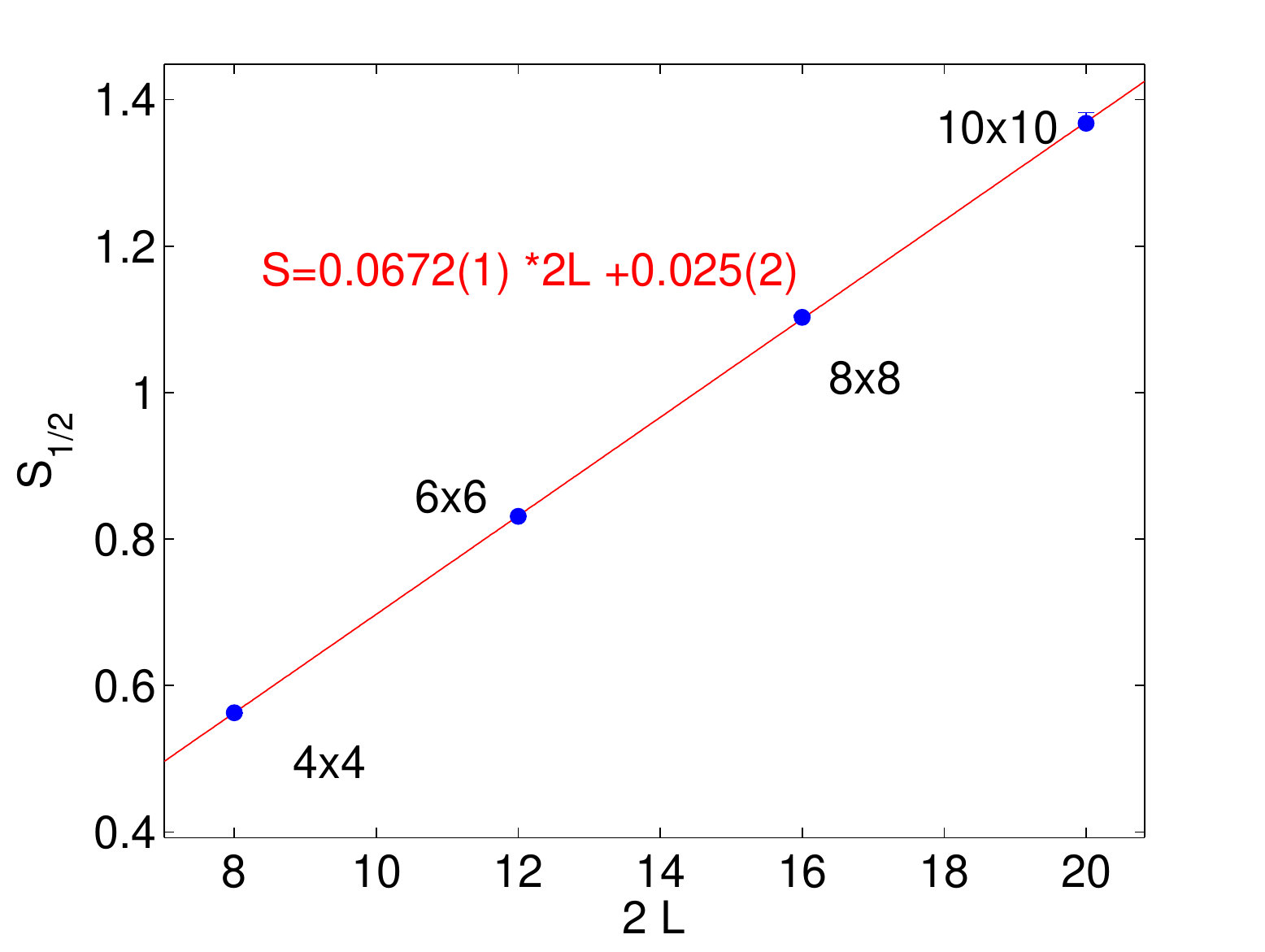}
\caption{Entropy $S_{1/2}(L)$ of one half of the torus as a function of the linear size $L$, corresponding to the ground state of $H_{\tmop{Ising}}$ with $\lambda = 3.044$. The results for $L={4,6,8,10}$ confirm the linear growth predicted in Eq. \ref{eq:EntHalf}, where no logarithmic correction is expected. The results of our study also seem to rule out the presence of the term proportional to $s_{-1}$. In the inset we show the results of the fit. The asymmetric errorbars, obtained through the analysis outlined in   section \ref{sec:error}, {are so small that they are hardly visible in the plot}. \label{fig:entropy}}
\end{center}
\end{figure}

 From the TTN approximation to the ground state of $H_{\tmop{Ising}}$ on a $L\times L$ torus for $L=\{4,6,8,10\}$ we can  compute the entanglement entropy of one half and one quarter  of the lattice at the critical point $\lambda=3.044$. Collecting results from  Refs. \cite{casini_entanglement_2005,casini_universal_2007,hsu_universal_2008,nishioka_holographic_2009,fradkin_entanglement_2006}, at a quantum critical point, the entanglement entropy  $S_{1/2}$  of half the torus, with total boundary $2 L$,   should scale  as
\begin{eqnarray}
   S_{1/2}(L) &=& s_1 2L +\frac{s_{-1}}{2L}  +\gamma_{QCP},
  \label{eq:EntHalf} 
  \end{eqnarray}
where $\gamma_{QCP}$ (see Ref. \cite{hsu_universal_2008, metlitski_entanglement_2009, fradkin_scaling_2009}) should be a universal constant.
 We  can now  numerically confirm the validity of  Eq. \ref{eq:EntHalf} and at the same time   extract  estimates for the coefficients in  Eq. \ref{eq:EntHalf}, including $\gamma_{QCP}$ for the Ising universality class on a torus.
This is  is presented in Fig. \ref{fig:entropy} where, by performing a fit to the numerical data with  Eq. \ref{eq:EntHalf},  we obtain  
\begin{eqnarray}
s_1 & = &0.06701(107),\\
\gamma_{QCP}& = &0.030(26),\\
s_{-1}    & =& -0.02(14),
\end{eqnarray}
 with  $\frac{\chi^2}{n.d.f.}= 0.003$ \footnote{In this section of the work the letter $\chi^2$ refers to the standard name given to the sum of the residual in a least square fit, and should not be confuse with the truncation parameter of the TTN  that is called $\chi$ through all the rest of work.}. The curve described by Eq. \ref{eq:EntHalf} is, hence, a good description of the scaling form of $S_{1/2}$. However the results we obtain are compatible with setting $s_{-1}$ to zero. If we do this, and repeat the fit we obtain
\begin{eqnarray}
\label{eq:num_ris_s}
s_1 & = & 0.06722(18) \\
\gamma_{QCP} & = &0.0250(21)\label{eq:num_ris_s1},
\end{eqnarray}
with $\frac{\chi^2}{n.d.f.}=0.002$ (slightly lower than the previous case). The values for $s_1$ and $\gamma_{QCP}$  are also compatible with the ones of the previous fit. Their accuracy is however  improved by one order of magnitude. 
These results  suggest the absence of the correction term $s_{-1}$ in the scaling of the entanglement entropy of half a torus.
It is interesting to notice that  $\gamma_{QCP} $, that  should be universal, is positive  as predicted in Ref. \cite{fradkin_scaling_2009}.

\begin{figure}[!tb]
\begin{center}
\includegraphics[width=8cm]{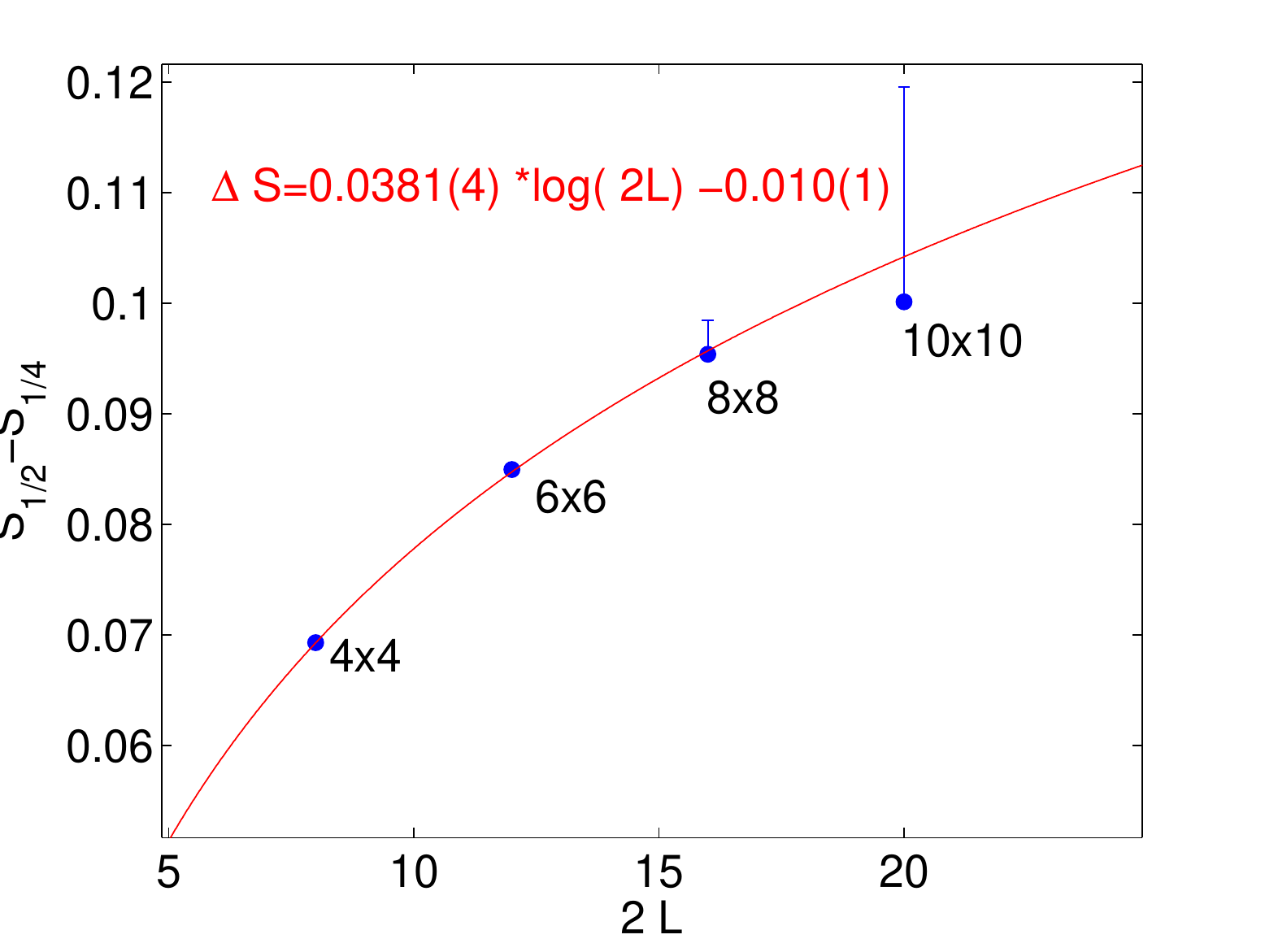}
\caption{Difference between entropies for one quarter and one half of the torus as a function of the linear size $L$, corresponding to the ground state of $H_{\tmop{Ising}}$ with $\lambda = 3.044$. The results for $L={4,6,8,10}$ allow us to confirm the logarithmic dependence predicted in Eq. \ref{eq:entro_log2}, which is attributed to the presence of corners in the boundary of our block for one quarter of the lattice.}
\label{fig:logarithm}
\end{center}
\end{figure}

Compiling results from  Refs. \cite{casini_entanglement_2005,casini_universal_2007,hsu_universal_2008,nishioka_holographic_2009, fradkin_entanglement_2006}, the entropy $S_{1/4}$ of one quarter of the torus  should scale  as
\begin{eqnarray}
   S_{1/4}(L) &=& s_1 2L +\frac{s_{-1}}{2L} +s_0 \log 2L + const.
 \label{eq:EntQuarter}
 \end{eqnarray}
where the presence of a logarithmic term is induced by the corners of the square block \cite{fradkin_entanglement_2006, casini_universal_2007, nishioka_holographic_2009, yu_entanglement_2008}. Comparing with Eq. \ref{eq:EntHalf} 
and using  that the interacting boundary is the same for one half and one quarter of the torus,  we can extract $s_0$  from 
\begin{equation}
S_{1/4}(L) - S_{1/2}(L) = s_0 \log 2L +const.
\label{eq:entro_log2}
\end{equation}
This study is presented in  Fig. \ref{fig:logarithm}. The fit to the numerical data with Eq.  \ref{eq:entro_log2}  produces an estimate 
\begin{equation}
\label{eq:num_ris_l}
 s_0 = -0.0381(4)  
\end{equation}
with a corresponding $\frac{\chi^2}{n.d.f.}=0.4$. This  confirms the validity of the scaling form of Eq. \ref{eq:EntQuarter} as well as a negative value for $s_0$ in agreement with the theory \cite{casini_entanglement_2005,fradkin_entanglement_2006,casini_universal_2007,nishioka_holographic_2009}. 

 As a side observation, this plot exemplifies once again the difference between the quasi-exact and the approximate regime. The level of approximation we obtain for the $L=10$ system implies that those results are almost useless in extracting logarithmic corrections to the leading scaling laws.

\subsection{Renyi entropies}
From the TTN approximation to the ground state, we can also compute all the Renyi entropies $S_n$
\begin{equation}
S_n= \frac{1}{1-n} log \tr \rho ^n \qquad 0\le n \le \infty
\end{equation}

 A limiting case  of $S_n$  is given by  the single copy entanglement $E^{(1)}$ \cite{eisert_single-copy_2005},
\begin{equation}
\label{eq:sing}
E^{(1)} \equiv \lim_{n\to \infty} S_n.
\end{equation}

\begin{figure}[!tb]
\begin{center}
\includegraphics[width=8cm]{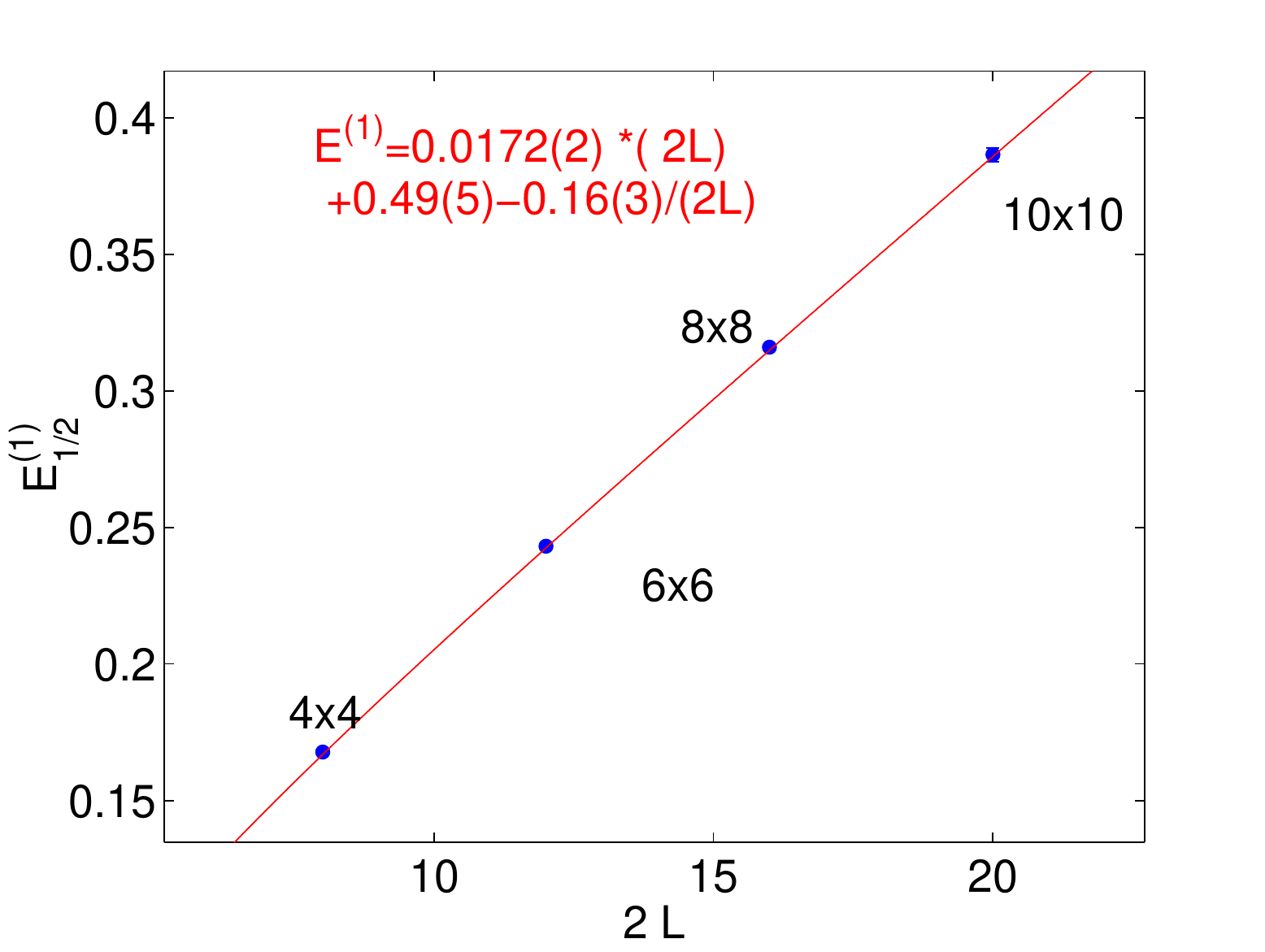}
\caption{Single Copy Entanglement $E^{(1)}(L)$ for  one half of the torus as a function of the linear size $L$, corresponding to the ground state of $H_{\tmop{Ising}}$ with $\lambda = 3.044$. The results for $L={4,6,8,10}$ confirm the linear growth predicted in Eq. \ref{eq:scal_sc}. In this case the term  $e_{-1}$, analogous to the term $s_{-1}$ for the entanglement entropy, is negative as predicted by the theory. In the inset  we present the results of the fit. The asymmetric errorbars, obtained through the analysis outlined in  section \ref{sec:error}, {are so small that they are hardly visible on the plot. }\label{fig:singlecopy}}
\end{center}
\end{figure}

$E^{(1)}$  is  expected to  have the same  scaling form of Eq. \ref{eq:EntHalf} for the entanglement entropy but  with different  numerical  coefficients \cite{stphan_shannon_2009,metlitski_entanglement_2009}, 
\begin{equation}
\label{eq:scal_sc}
 E^{(1)}_{1/2}(L) =e_1 2L +\frac{e_{-1}}{2L} + \gamma'_{QCP} , 
\end{equation}
The results for $E_1$ of half torus are shown in \ref{fig:singlecopy}. The fit to the numerical data for $L={4,6,8,10}$ with Eq. \ref{eq:scal_sc} produces

\begin{eqnarray}
\label{eq:num_ris_e}
e_1 & = & 0.01724(20),\\ 
\gamma'_{QCP}& = & 0.0499(51), \label{eq:num_ris_e2}\\
e_{-1}    & =& -0.161(30)\label{eq:num_ris_e3},
\end{eqnarray}
with $\frac{\chi^2}{n.d.f.}=0.0005$. This reveals a very good agreement between Eq. \ref{eq:scal_sc} and the numerical data.
It is interesting to notice that   the coefficient  $e_{-1}$ is non-zero and  negative, in agreement with the theory \cite{casini_entanglement_2009}.
In addition, as already anticipated by the results contained in Ref. \cite{metlitski_entanglement_2009,stphan_shannon_2009}, the numerical values of the parameters for scaling of the single copy entanglement and entanglement  entropy are different. Nevertheless, for the universal term, we find that 
\begin{equation}
\label{eq:sing-ent}
\gamma'_{QCP}= 2 \gamma_{QCP} 
\end{equation} 
to our numerical precision.
This is reminiscent of a very similar result obtained for one dimensional critical chains, where the universal coefficient of the logarithmic scaling of the entanglement entropy with the size of the interval is two times (instead than one half as in Eq. \ref{eq:sing-ent} ) the analogous  coefficient  for  the scaling of the single copy entanglement \cite{orus_halfentanglement_2006,peschel_single-copy_2005}.

\section{Discussion}
\label{sect:discussion}
In this manuscript we have described a numerical technique based on a TTN  to compute ground state properties of 2D lattice systems. The approach exploits  the entropic area law and has a cost that scales exponentially in the linear size of the lattice. Its goals are necessarily more modest than those of scalable tensor network algorithms such as  PEPS and MERA \cite{sierra_density_1998,maeshima_vertical_2001,nishio_tensor_2004,gu_tensor-entanglement_2008,verstraete_renormalization_2004,murg_variational_2007,jordan_classical_2008,murg_exploring_2009,vidal_entanglement_2007,vidal_class_2006,evenbly_entanglement_2008}.

 For the model we have considered here, the TTN approach offers  a simple, effective way of obtaining quasi-exact results well beyond what is possible with exact diagonalisation techniques  \cite{hamer_finite-size_2000,henkel_statistical_1984}. We expect that similar gains would also occur for other models.  We envisage that this technique will become a useful tool both to study small lattice systems and in investigations based on finite size scaling. A highlight of the approach is its simplicity, specially when compared to the scalable tensor network algorithms. In addition, it can be used to study block entropies, a task that becomes much less straightforward with other methods.

The TTN algorithm is closely related to the DMRG algorithm applied to 2D lattices. It is beyond the scope of the present work to conduct the detailed analysis required to establish how the performances of the two algorithms compare. Nevertheless, some preliminary observations can be made. Updating the matrix product state (MPS) used in DMRG has a cost of $O(\chi^3L^2)$ per sweep, while updating the TTN costs $O(\chi^4L)$. This allows DMRG to consider values of $\chi$ that are about 10 times larger with similar computational cost. On the other hand the TTN has  better connectivity. In a TTN all lattice sites are connected through the product of at most $O(\log L)$ tensors. Instead, when an MPS is used to encode the ground state of a 2D lattice, nearest neighbour lattice sites are typically connected through the product of $O(L)$ tensors, with a fraction of the sites being connected through the product of $O(L^2)$ tensors (on a torus). As a result, we expect convergence to the ground state to be faster using a TTN. In addition, space symmetries can be (partially) incorporated in a TTN.

The TTN is particularly fitted to study entropies and their scaling with the size of the system. In this work we have reported some  numerical results that are compatible with the expectation drawn from Refs. \cite{casini_entanglement_2005,fradkin_entanglement_2006,casini_universal_2007,hsu_universal_2008} about the presence, in the scaling form of the entanglement entropy, of both additive logarithmic and constant  corrections to the area law. Our results suggest the absence of a $1/L$ correction. After the first draft of our paper was presented, a systematic study of all the Renyi entropies was reported   in Ref. \cite{metlitski_entanglement_2009}. This motivated us to also  consider the single copy entanglement $E^{(1)}$, Eq. \ref{eq:sing}. We have confirmed that the  scaling of $E^{(1)}$ also includes a constant  additive correction to the area law and a term proportional to $1/L$, with numerical values different from the ones present in the scaling of the entanglement entropy. This is a  hint to the presence of a different set of universal constants for each of the Renyi entropies as stated in Ref. \cite{metlitski_entanglement_2009}. Further aspects of this scenarios can be found in Ref. \cite{gliozzi_entanglement_2009}.

The authors thank Philippe Corboz, Ian McCulloch, Robert Pfeifer,  for useful conversations and technical advice. We also  would like to thank F. Gliozzi, M. Huerta and C. Hamer for useful correspondence. Financial support from the Australian research council (APA, FF0668731, DP0878830) is acknowledged.







\end{document}